\documentclass[review]{elsarticle}

\usepackage{mathptmx}       
\usepackage{helvet}         
\usepackage{courier}        
\usepackage{type1cm}        
\usepackage{makeidx}         
\usepackage{graphicx}        
\graphicspath{../figures/}
\DeclareGraphicsExtensions{.pdf,.jpg,.jpeg,.png,.eps}
\usepackage{multicol}        
\usepackage[bottom]{footmisc}
\usepackage{amssymb}
\usepackage{amsmath}
\usepackage{epsfig}

\usepackage{amsmath}
\usepackage{amsfonts}
\usepackage{amssymb}

\usepackage{tabularx}

\newcolumntype{L}[1]{>{\raggedright\arraybackslash}p{#1}}
\newcolumntype{C}[1]{>{\centering\arraybackslash}p{#1}}
\newcolumntype{R}[1]{>{\raggedleft\arraybackslash}p{#1}}

\begin{document}

\begin{frontmatter}

\title{Modelling the Demand and Uncertainty of Low Voltage Networks and the Effect of non-Domestic Consumers}


\author[label1]{Georgios Giasemidis}
\address[label1]{CountingLab LTD and CMoHB, University of Reading, RG6 6AX, Reading, UK}
\ead{g.giasemidis@reading.ac.uk}

\author[label2]{Stephen Haben}
\address[label2]{Mathematical Institute, University of Oxford, OX2 6GG, Oxford, UK}
\ead{Stephen.Haben@maths.ox.ac.uk}


\begin{keyword}
	low voltage networks \sep load demand modelling \sep genetic algorithm \sep 
	buddying \sep uncertainty \sep confidence bounds \sep commercial customers.
\end{keyword}

\begin{abstract}
The increasing use and spread of low carbon technologies are expected to cause new patterns in electric demand and set novel challenges to a distribution network operator (DNO). 
In this study, we build upon a recently introduced method, called ``buddying'', which simulates low voltage (LV) networks of both residential and non-domestic (e.g. shops, offices, schools, hospitals, etc.) customers through optimization (via a genetic algorithm) of demands based on limited monitored and customer data. 
The algorithm assigns a limited but diverse number of monitored households (the ``buddies'') to the unmonitored customers on a network. 
We study and compare two algorithms, one where substation monitoring data is available and a second where no substation information is used. 
Despite the roll out of monitoring equipment at domestic properties and/or substations, less data is available for commercial customers. 
This study focuses on substations with commercial customers most of which have no monitored ``buddy'', in which case a profile must be created. 
Due to the volatile nature of the low voltage networks, uncertainty bounds are crucial for operational purposes. 
We introduce and demonstrate two techniques for modelling the confidence bounds on the modelled LV networks. 
The first method uses probabilistic forecast methods based on substation monitoring; the second only uses a simple bootstrap of the sample of monitored customers but has the advantage of not requiring monitoring at the substation. 
These modelling tools, buddying and uncertainty bounds, can give further insight to a DNO to better plan and manage the network when limited information is available.
\end{abstract}
\end{frontmatter}


\section{Introduction}
\label{sec:intro}

Modern environmental policies to reduce carbon emissions are leading many countries towards a low carbon economy.
The increased interest and use of relatively new technologies, such as photovoltaics, electric vehicles, storage devices, etc., drastically change the way customers use and generate their energy, setting novel challenges to distribution network operators (DNO) to manage and plan the networks. Accurate modelling of the individual customers and their aggregate demand on the network is crucial for assessing and anticipating the stability and headroom on the LV network \cite{Harrison2005}. This situation has sparked the interest of both the academic and DNO communities in modelling LV networks, as traditional approaches might be inadequate. 
Some of these techniques are based on clustering smart meters \cite{Flath2012,Stephen2014,McLoughlin2015}, resulting in smoothed/averaged profiles which are not representative of the volatile profiles of individuals. 
Others adopt a bottom-up methodology \cite{Richardson2010,Capasso1994,Muratori2013465}, where a customer’s profile is created by aggregating the demand of individuals’ appliances, or other combine smart meter data with individual surveys \cite{Kipping2015655}.
However, these techniques require granular, complex data, and understanding of behavioural patterns of the customers, which are usually obtained by questionnaires and are not frequently updated \cite{Muratori2013465}. 

Across the world, smart meters are being rolled out, enabling monitoring of large parts of the networks. 
Customers’ profiles are valuable for network operators as these profiles can be used in network modelling environment tools to perform power flow and voltage analysis \cite{Cyme}. 
However, particularly in the UK, DNOs do not have free access to smart meters, as they are owned by the utility companies, and acquiring large-scale monitoring data might be expensive. 
DNOs, though, might have access to aggregated data, such as substation/feeder monitoring data or quarterly meter readings (QMR) of the customers.

In this study, we present a method for accurately populating a LV network of unmonitored customers, both domestic and non-domestic, with realistic profiles from a sample of monitoring customers, which are assumed to be diverse and representative of the population. 
In particular, we assign every customer on the network to a monitored profile using a technique called buddying, which was recently introduced by the authors in \cite{Giasemidis2016} and extended in this study to accommodate non-domestic customers too.
This technique is split into two methods, depending on whether substation (i.e. aggregate) level monitoring is available or not.

As LV networks have volatile profiles, modelling a single profile of a substation or feeder might not give a full description of the capacity of the network. 
Uncertainty intervals are essential to ensure that the diversity is incorporated. Hence, we also introduce methods for estimating the uncertainty of LV networks. 
Similar to the buddying methods, we introduce two methods for estimating the uncertainty, one that requires substation data and another that does not.

Our methods can be combined with LCTs, e.g. solar panels, either explicitly or implicitly. First, the data-set of monitored households includes 28 properties with solar panels too, which are used to model (``buddy'') the customers with solar panels in the studied network (there is 0.7\% penetration of solar panels in the studied network). Additionally, export and import profiles can be used to model the energy generation of the solar panels and also fed into the DNO's network tools to assess the stresses on the network. The method can also be easily generalised to accommodate properties with other LCTs if they exist in a network. In this case, the dataset of monitored customers should be expanded to include a sample of customers that have adopted the particular technology. 
An implicit utility of the proposed methodology is that it creates baseline demand profiles for all customers on a network. These profiles have been used with uptake scenarios of LCTs to assess the future stresses on a network \cite{Hattam2017,HATTAM2018353}.

The novel aspects of this study are two-fold. First, we extend the buddying method introduced in \cite{Giasemidis2016} to non-domestic customers. This is a non-trivial task, as there are no monitored profiles of non-domestic customers in the dataset. For this reason, profiles of non-domestic customers must be generated from profile tools and their modelling must account for the diversity across types of customers. Second, methods for estimating confidence bounds (or uncertainty intervals) are also introduced, which can be used with or without the results from the buddying method (depending whether substation monitoring is available or not).

The rest of the paper is organised as follows. In Section \ref{sec:buddying}, we introduce the buddying methodology in \cite{Giasemidis2016} and extend it to non-domestic customers which lack monitored profiles. Section \ref{sec:nondomestic-feeders} presents their characteristics and modelling challenges. Section \ref{sec:confidence} introduces and assesses techniques for estimating the uncertainty of a LV network. Finally, we summarise the results of this paper in Section \ref{sec:conclusion}.

\section{Buddying}
\label{sec:buddying}

\subsection{Methodology}
\label{sec:buddying-methodology}

In the absence of smart meter data for every customer on the LV network, buddying is used to simulate the demand profiles of each customer on a feeder. Let a LV feeder have $M$ connected customers, labelled $c_j, j = 1, \ldots, M$. These customers have no smart meter data available, but their mean daily demand $U_j, j = 1, \ldots, M$ is known from their quarterly meter readings, which are typically available to network operators. Additionally, we assume access to a sample of $N$ half-hourly profiles of customers through smart meters. On a typical network, the total number of unmonitored customers is significantly larger than the sample of monitored profiles, i.e. $M_{\text{total}} >> N$. The monitored profiles span a period of $d$ days and we denote this set as
\begin{equation}
	\mathcal{P} = \{\mathbf{p}_k = \left(p_k(1), \ldots, p_k(48 d)\right)^T \in \mathbb{R}^{48d} \quad | \quad k = 1, \ldots, N\}.
	\label{eq:monitors-set}
\end{equation}

The mean daily demand of the monitored profiles is found directly from their profiles, i.e.
\begin{equation}
	\hat{U}_k= \frac{1}{d} \sum_{h=1}^{48d}p_k(h).
	\label{eq:mdd} 
\end{equation}

Network operators might have access to substation/feeder monitoring data or, alternatively, they may have access to the aggregated demand of all customers connected on the feeder. We denote this profile by $\mathbf{s} = \left(s(1), \ldots, s(48d)\right)^T \in \mathbb{R}^{48d}$.

The proposed method aims to assign a profile $\mathbf{p}_k \in \mathcal{P}$  from the sample of monitored profiles to every unmonitored customer, $c_j$, on a LV network. This method is called buddying and the profile $\mathbf{p}_k$ is referred to as the buddy of customer $c_j$. Note that different unmonitored customers might have the same buddy. 

We also split all customers, both monitored and unmonitored, into groups based on the available demographics information. Buddying is designed to select buddies that belong to the same group in order to reduce the search space of the parameters. Here, we use the profile class and council tax band%
\footnote{Council Tax is a local taxation system used in Great Britain on domestic properties. Each property is assigned one of eight bands (A to H) approximately based on property value.} 
information. In the UK, customers are assigned to ELEXON profile classes (PC) \cite{Elexon2013}. There are eight generic profile classes representative of large populations of similar customers. Classes 1 and 2 correspond to domestic customers and distinguish between two tariffs, ``Standard'' and ``Economy 7''. The latter provides cheaper rates overnight at the expense of increased day-time charges.
Domestic customers with profile classes 1 and 2 are split into 6 groups, based on their council tax band, 
whereas non-domestic customers are grouped according to types, e.g. hospitals, schools, offices, restaurants, supermarkets, etc. For further details on the groups, please see \cite{Giasemidis2016}.

Smart meter data represents only domestic customers. No monitored profiles for non-domestic customers are currently available for this study. However, we have standard profiles for each particular type of non-domestic customer, which are part of current profile tools (such as WinDEBUT \cite{Windebut}). This standard profile is normalised so that the daily mean usage is 1kWh and is part of the monitored set $\mathcal{P}$ in Eq. \eqref{eq:monitors-set}.

Here, we present two algorithms for buddying, a simple algorithm (SA), which assumes that no substation monitoring is available and an approach based on a genetic algorithm (GA), which requires substation/feeder monitoring data.

\subsubsection{Simple Algorithm}
\label{sec:sa}
For domestic properties, the simple algorithm assigns the profile $\mathbf{p}_k \in \mathcal{P}$ which is in the same group as the customer $c_j$ and has the closest mean daily usage, i.e. 
\begin{equation}
	k = \arg\min_{i \in I_g} |U_j - \hat{U}_i|,
	\label{eq:sa}
\end{equation}
where $I_g$ is the index for profiles in group $g$. For non-domestic properties, the algorithm simply scales the normalised standard profiles by their mean daily demand. This method does not use the substation data.

\subsubsection{Genetic Algorithm}
\label{sec:ga}

If the substation/feeder monitoring data is available, we develop more sophisticated techniques for buddying. The aim of the buddying is to find a set of profiles $\hat{\mathcal{P}} = \{\mathbf{p}_{k_1}, \ldots, \mathbf{p}_{k_M}\}$, where $\mathbf{p}_{k_j} \in \mathcal{P}$ is the buddy for unmonitored customer $c_j$ for $j = 1, \ldots, M$, that minimises the cost function
\begin{equation}
	F(\hat{\mathcal{P}}) = (1-w) \sum_{h=1}^{48d}\frac{|a(h) - s(h)|}{S} + w \left(\sum_{j=1}^{M_{\text{dom}}} \frac{|U_j - \hat{U}_{k_j} |}{D} + \sum_{j=M_{\text{dom}}+1}^{M_{\text{com}}} \frac{U_j |(1 - \alpha_j)|}{D}\right),
	\label{eq:cost-function}
\end{equation}
over all possible sets of buddied profiles $\hat{\mathcal{P}}$. Here, $M_{\textbf{dom}}$ and $M_{\textbf{com}}$ are the number of domestic and non-domestic (usually commercial) properties, respectively, on the feeder ($M = M_{\textbf{dom}} + M_{\textbf{com}}$). $\alpha \in [0.8, 1.2]$ is a scaling parameter that is randomly sampled from a uniform distribution and controls the scaling of the normalised standard profiles. $a(h)$ represents the aggregate demand of the buddy, i.e.
\[a(h) = \sum_{j=1}^{M_{\text{dom}}} \mathbf{p}_{k_j}(h) + \sum_{j=M_{\text{dom}}+1}^{M_{\text{com}}} \alpha_j U_j \mathbf{p}_{k_j}(h). \]
Finally, $S = \sum_{h=1}^{48d} s(h)$ and $D = \sum_{j=1}^M U_j$.

The first term in Eq. \eqref{eq:cost-function} controls the fit of the aggregated profiles to the feeder readings, whereas the second term controls the buddy based on the individuals’ mean daily demand. The weighting $w \in [0,1]$ allows the buddy to either optimise fully to the substation ($w=0$) or completely to the mean daily usages ($w=1$), or a weighted sum of the two. The optimal choice in $w$ can also be interpreted as the trust in the accuracy of the quarterly meter readings (since they are generally estimates) as well as their importance in identifying an accurate buddying. The case with $w=1$ uses no substation data and is equivalent to the SA.

The optimal collection of buddies is found by implementing a genetic algorithm where the fitness function is Eq. \eqref{eq:cost-function}. For further details of the GA and its implementation, see \cite{Giasemidis2016}.

The SA buddy is expected to be less accurate at assigning buddying profiles than the GA buddy, because no substation data is used. However, this option is more attractive to a network operator since it uses the minimal amount of monitoring and hence reduces potential data costs and storage.

The quality of the buddying can be assessed at both the aggregate level, i.e. how well the aggregate profile matches the substation feeders, and at the individual level, i.e. how accurately a buddied profile matches a monitored profile. Since, we have only a few monitored profiles on the network under study, assessing the buddy at the individual level is not possible. Further research in this area using pseudo-feeders has been conducted in \cite{Giasemidis2016} and shows that, at least for domestic-only feeders, the buddying is accurate at the customer level. Here, we focus on the substation accuracy and consider the relative mean absolute error (RMAE), defined by
\begin{equation}
	RMAE = \frac{1}{S} \sum_{t=1}^{T} | s(t) - a(t) |,
	\label{eq:rmae}
\end{equation}
where $s$ is the actual profile (either individual domestic household or substation) and $a$ is the estimated profile (either the buddy or the aggregate of the buddies).

\subsection{Data Description}
\label{sec:dataDescription}

Our dataset consists of half hourly energy data between 20th March 2014 and 22nd September 2015 inclusive from 54 LV substations corresponding to 191 feeders (in Bracknell, UK), collected as part of the Thames Valley Vision project\footnote{http://www.thamesvalleyvision.co.uk/}. 
There are about 8,000 customers connected to this network. 
We also use 242 monitored domestic profiles at half-hourly resolution for the same trial period and 29 different types of non-domestic customers with standardised profiles produced from the WinDEBUT tool \cite{Windebut}, see next paragraphs for further details. We pre-processed the raw data to replace missing values, outliers and anomalous readings with the average load from similar hours. We use the learning outcomes of \cite{Giasemidis2016}  on seasonality and length of training and use a period of 8 weeks starting on 5th of January 2015 for training. We also explore how the weighting parameter affects buddying using a range of weighting parameters from 0 to 1 in increments of 0.1. The test period is chosen to be an entire year (to reduce biases due to seasonal effects) from 01/09/2014 until 31/08/2015.

From 191 feeders, 127 are connected to domestic customers only (these feeders will be called domestic-only feeders), 58 have both domestic and non-domestic properties and will be named non-domestic or commercial feeders, whereas 6 feeders have only non-domestic properties and will be referred to as purely non-domestic or purely commercial feeders. We elaborate on the latter type of feeders in Section \ref{sec:nondomestic-feeders}.

\begin{figure}
	\centering
	\includegraphics[scale=0.21]{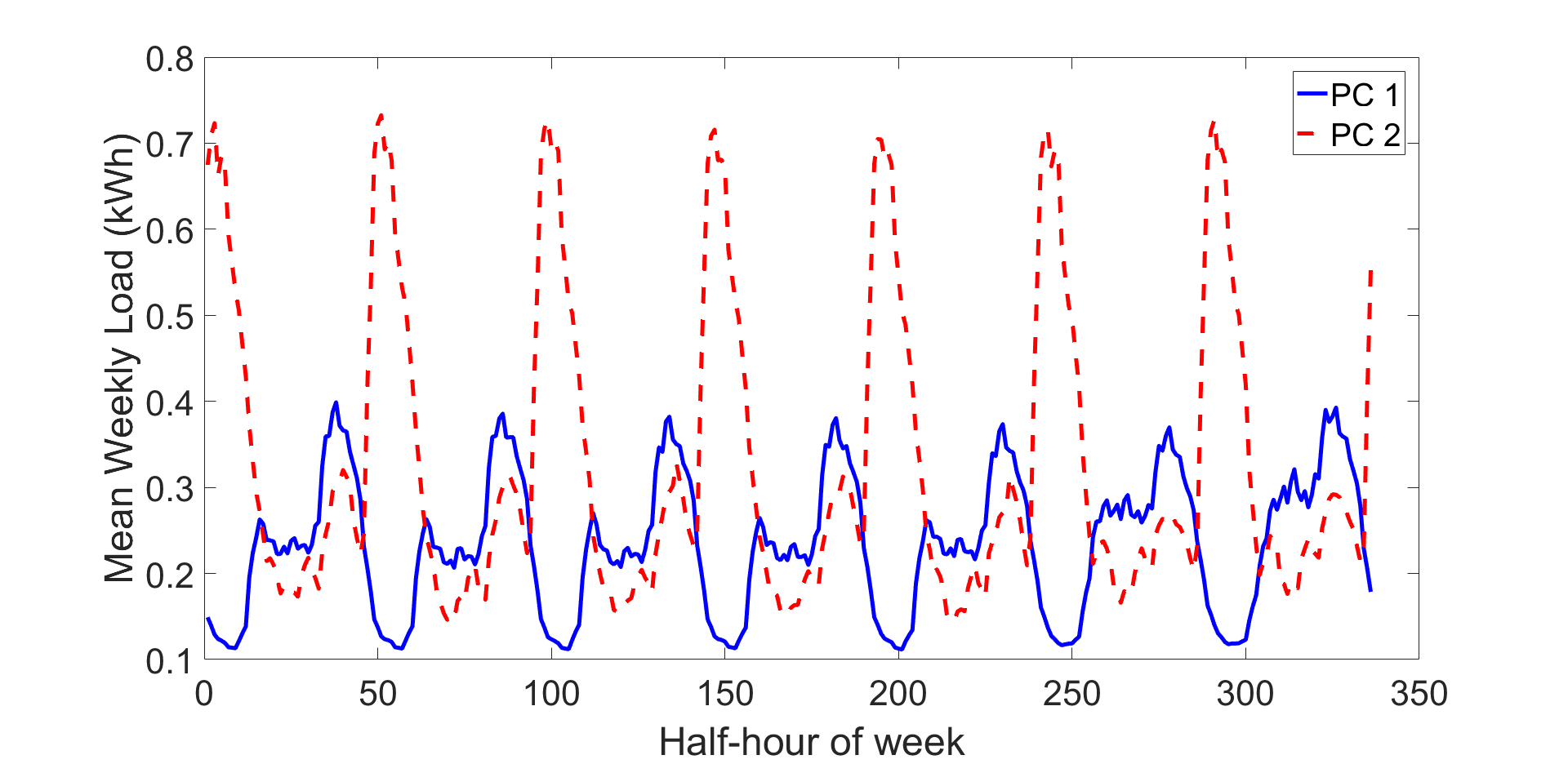}
	\caption{Mean weekly demand of the average profile class 1 (blue) and 2 (red) customer.}
	\label{fig:mean-demand-pc1-pc2}       
\end{figure}

The domestic smart meters are distinguished between PC 1 and PC 2, with or without solar panels. The two profile class types have distinguished profiles, see Figure \ref{fig:mean-demand-pc1-pc2}, where we plot the mean normalised weekly demand of each profile class. From Figure \ref{fig:mean-demand-pc1-pc2}, we observe that PC 2 customers have their peaks at night-hours (due to overnight storage heaters), when PC 1 customers have the lowest load values.

In Figure \ref{fig:mean-demand-pc1-pc2}, we excluded smart meters with photovoltaics, since their day-time demand is altered due to generation of energy. Table \ref{tbl:pc1-2-summary} shows the set size, the minimum, maximum, mean and standard deviation of the mean daily demand (kWh) of the two sets of customers, PC 1 and PC 2. We observe that customers in PC 2 have higher mean daily demand, as observed from Figure \ref{fig:mean-demand-pc1-pc2} too, and greater demand variability.

The set of monitored domestic profiles is diverse and representative of the entire network  under consideration (Bracknell area, UK) for the following reasons. First, it covers properties with all available council tax bands, profile classes and their combinations that appear in the network. These two demographic characteristics are indicative of behavioural patterns, as shown in Figure \ref{tbl:pc1-2-summary}. Our dataset also includes 28 monitored customers with photovoltaics, which are representative of properties with solar panels that exist in the network. In addition, these monitored properties span a wide range of mean demands as shows in Table \ref{tbl:pc1-2-summary}. Furthermore, the monitored profiles come from the same area as the LV network under consideration, they have therefore consistent response to the weather effects, and hence they are representative of the entire population of the studied network. Increasing the number of monitored properties does not necessarily guarantee better results for the algorithm, as the genetic algorithm will search in a higher dimensional parameter space, which is both computationally expensive and it might result to local minima that are away from the global minimum. Finding lower and upper bounds on the number of monitored customers for the algorithms to be efficient is beyond the scope of this work. Finally, if a new type of domestic customer is connected to the network (e.g. with a new type of LCT), monitored customers of this new type must be added to the smart-meter set.

\begin{table}[]
	\centering
	\begin{tabular}{|l|R{1.7cm}|r|r|r|r|}
		\hline
		& Number of Customers  & Min (kWh) & Max (kwh) & Mean (kWh) & Std (kWh) \\
		\hline  \hline 
		PC 1 & 199 & 0.761     & 35.775    & 11.245     & 6.427     \\
		\hline
		PC 2 & 15  & 4.800     & 46.110    & 16.015     & 11.678    \\
		\hline      
	\end{tabular}
	\caption{Summary statistics of the mean daily demand of domestic smart meter profiles of domestic properties without photovoltaics.}
	\label{tbl:pc1-2-summary}
\end{table}

The yearly profiles for the non-domestic are generated from a typical normalised weekly profile. The weekly profiles are produced using the profile tool WinDEBUT \cite{Windebut}. The profiles are extracted and processed to create a ``typical'' profile for the commercial customer and then normalised so that their mean daily usage is 1 kWh. Holidays and special days are replaced with the non-operational days from the typical weekly profile. For example, the weekend profile is used for the summer period for a school profile. We discuss the development of commercial profiles in more detail in Section \ref{sec:nondomestic-feeders}.

\subsection{Results}
\label{sec:results}

\begin{figure}
	\centering
	\includegraphics[width=\textwidth]{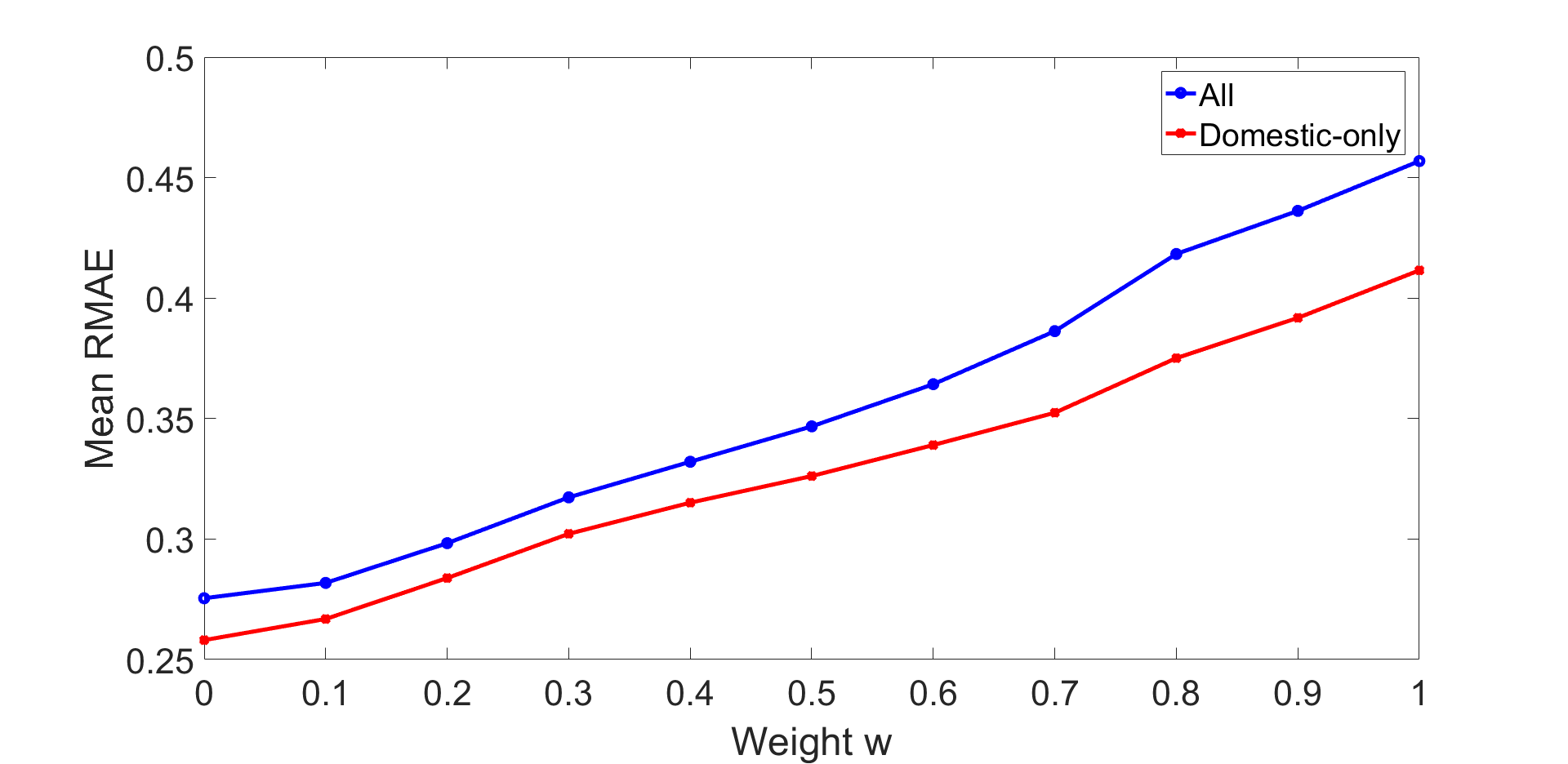}
	\caption{Average RMAE errors of all (blue) and domestic-only feeders (red) as a function of the weight parameter $w$.}
	\label{fig:rmae-vs-w}       
\end{figure}

Here, we present the error analysis of the domestic and non-domestic feeders. For an in-depth analysis on domestic-only feeders, see \cite{Giasemidis2016}. Figure \ref{fig:rmae-vs-w} shows how the mean RMAE (Eq. \eqref{eq:rmae}) varies with the weight parameter when considering all (blue) and domestic-only (red) feeders. We observe that the average error increases as the weight increases. This is consistent with our expectations and the findings in \cite{Giasemidis2016}, because the error score Eq. \eqref{eq:rmae} favours the buddying with zero weight, i.e. training the buddying on feeder readings only.

From Figure \ref{fig:rmae-vs-w}, we also observe that the commercial feeders are less accurately modelled compared to the domestic-only feeders. Our assumptions on the standard normalised profiles and the way we scale them require further refinements, see Section \ref{sec:nondomestic-feeders} for a further discussion on non-domestic feeders. A common problem is that the mean daily demand of many non-domestic properties is not known. In this case, it is assigned the average of the mean daily demands of the same type of properties. This results in inaccurately scaled non-domestic profiles.

\begin{figure}
	\centering
	\includegraphics[scale=0.3]{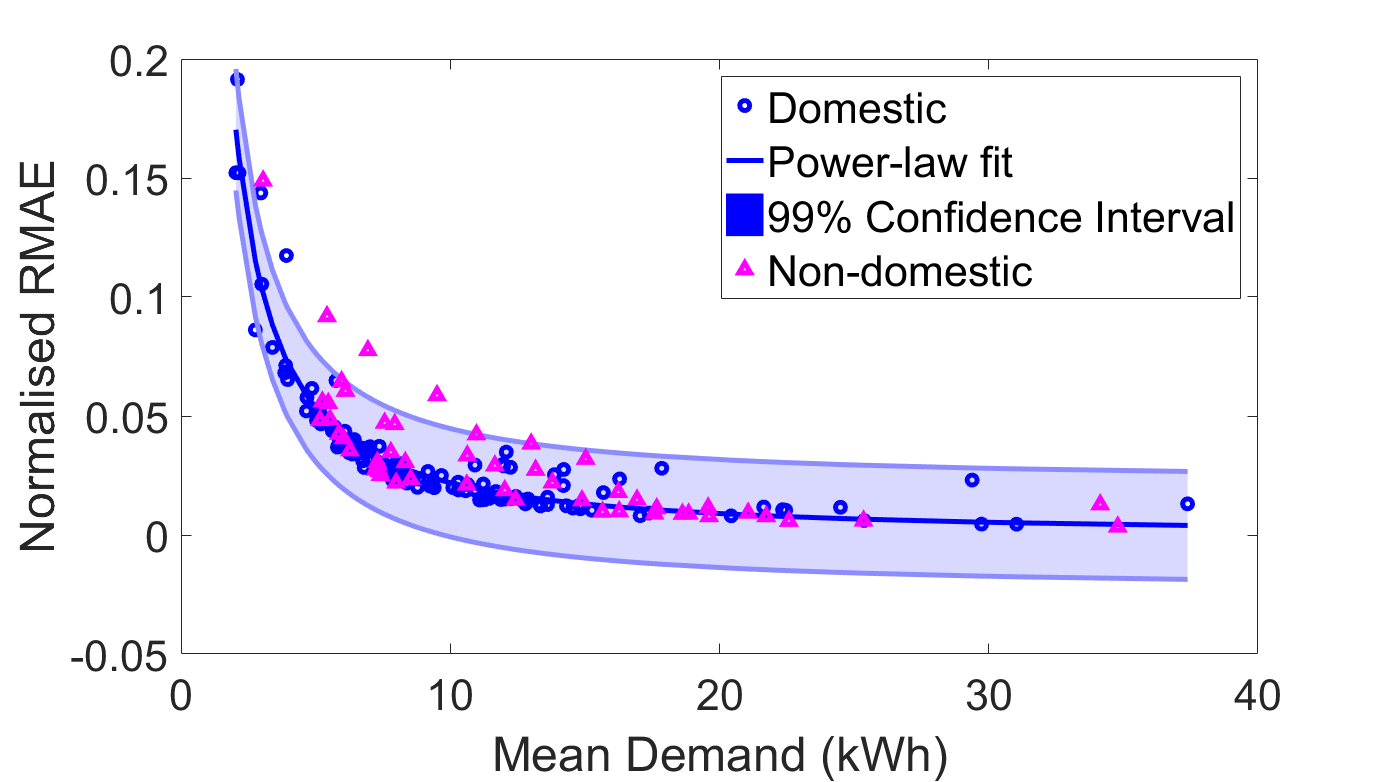}
	\caption{RMAE for the GA ($w=0$) method as a function of the feeder's mean demand. The fit was applied to domestic-only feeders (blue).}
	\label{fig:rmae-vs-meandemand-ga}       
\end{figure}

An important question to a DNO is how the errors are distributed, and whether the errors are correlated to the other feeder characteristics. In Figure 3, we plot the feeder's error score RMAE (Eq. \eqref{eq:rmae}) divided by its mean demand as a function of the mean demand of the feeder for the GA case with $w = 0$. We apply a power law fit of the form $a \cdot x^{-b}$, for positive $a, b$, considering only the domestic feeders, an established result from \cite{Giasemidis2016}. Hence, using this curve, and its confidence bar, we can thus estimate the size of buddying error, whatever the mean demand of the feeder. In Figure \ref{fig:rmae-vs-meandemand-ga}, we also plot the non-domestic feeders and observe that the errors, in general, fit within the 99\% confidence bounds of the power-law fit, with only four non-domestic feeders lying outside. Additionally, for feeders with mean demand greater than 15 kWh, the error is close to the fit. Similar to the observations in \cite{Giasemidis2016}, there is a region of mean demand, between 5 to 15 kWh, for which some of the errors vary significantly from the fit.

\begin{figure}
	\centering
	\includegraphics[scale=0.3]{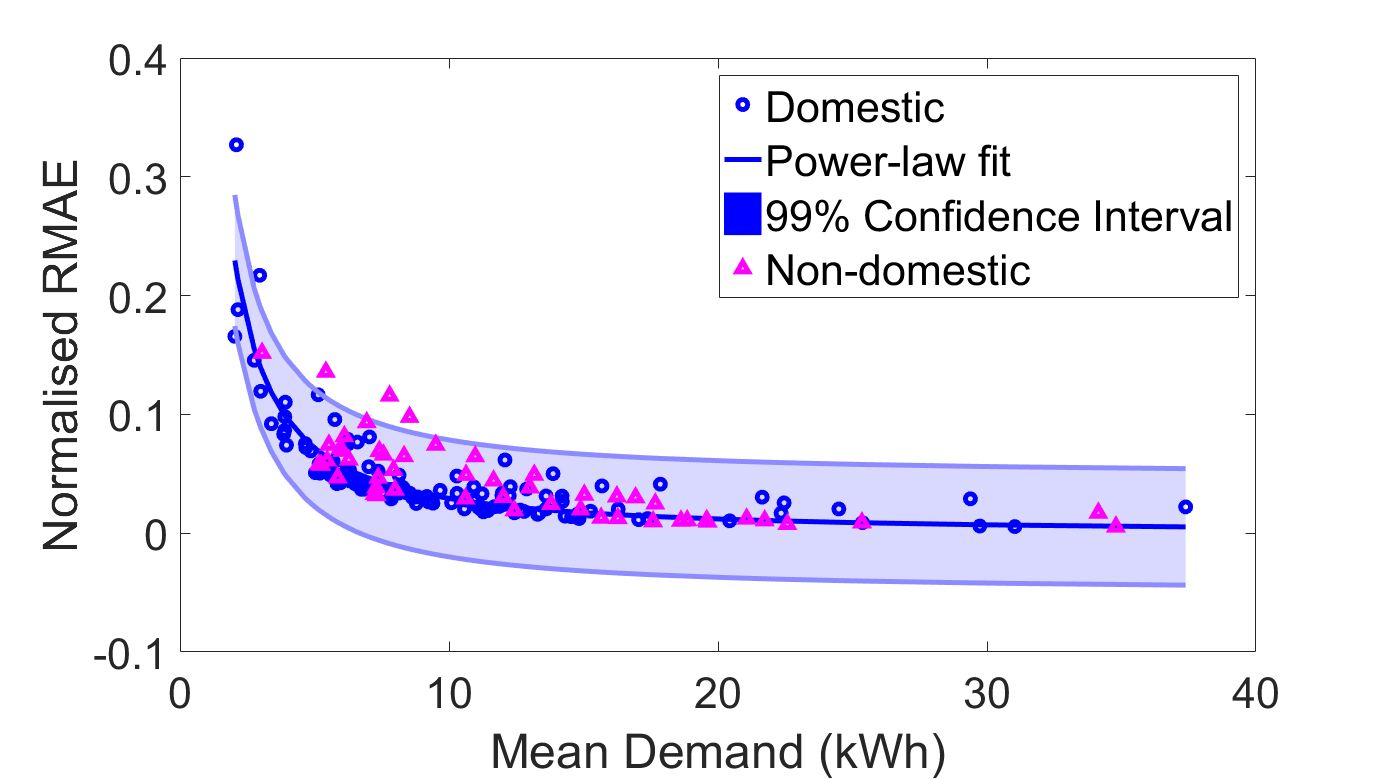}
	\caption{RMAE for the SA method as a function of the feeder's mean demand. The fit was applied to domestic-only feeders (blue).}
	\label{fig:rmae-vs-meandemand-sa}       
\end{figure}

A similar plot to Figure \ref{fig:rmae-vs-meandemand-ga} is presented in Figure \ref{fig:rmae-vs-meandemand-sa} for the SA method.
We notice that the SA errors are larger than the GA ones. The SA errors are more variable for fixed mean demand, whereas the GA errors follow a power-law pattern more closely. The GA errors indicate that the buddying accuracy is greater for feeders with a higher mean demand. From a DNO perspective feeders with larger mean demand are more accurately modelled and thus potentially reduces the need for monitoring. On the other hand, feeders with low demand have the lowest buddying accuracy, but they have more headroom, hence no monitoring is needed. It is feeders that have a significant total demand (usually supplying a few tens of customers), and relative high errors that might need further monitoring. Thus, the power-law fit can guide DNOs about the expected modelling capabilities for a LV network of a certain size, and whether monitoring is required.

The power-law scaling behaviour is consistent to existing studies in the academic literature \cite{Sevlian2014}, where the authors studied the effect of aggregation on short term load forecasting.

\begin{figure}
	\centering
	\includegraphics[scale=0.2]{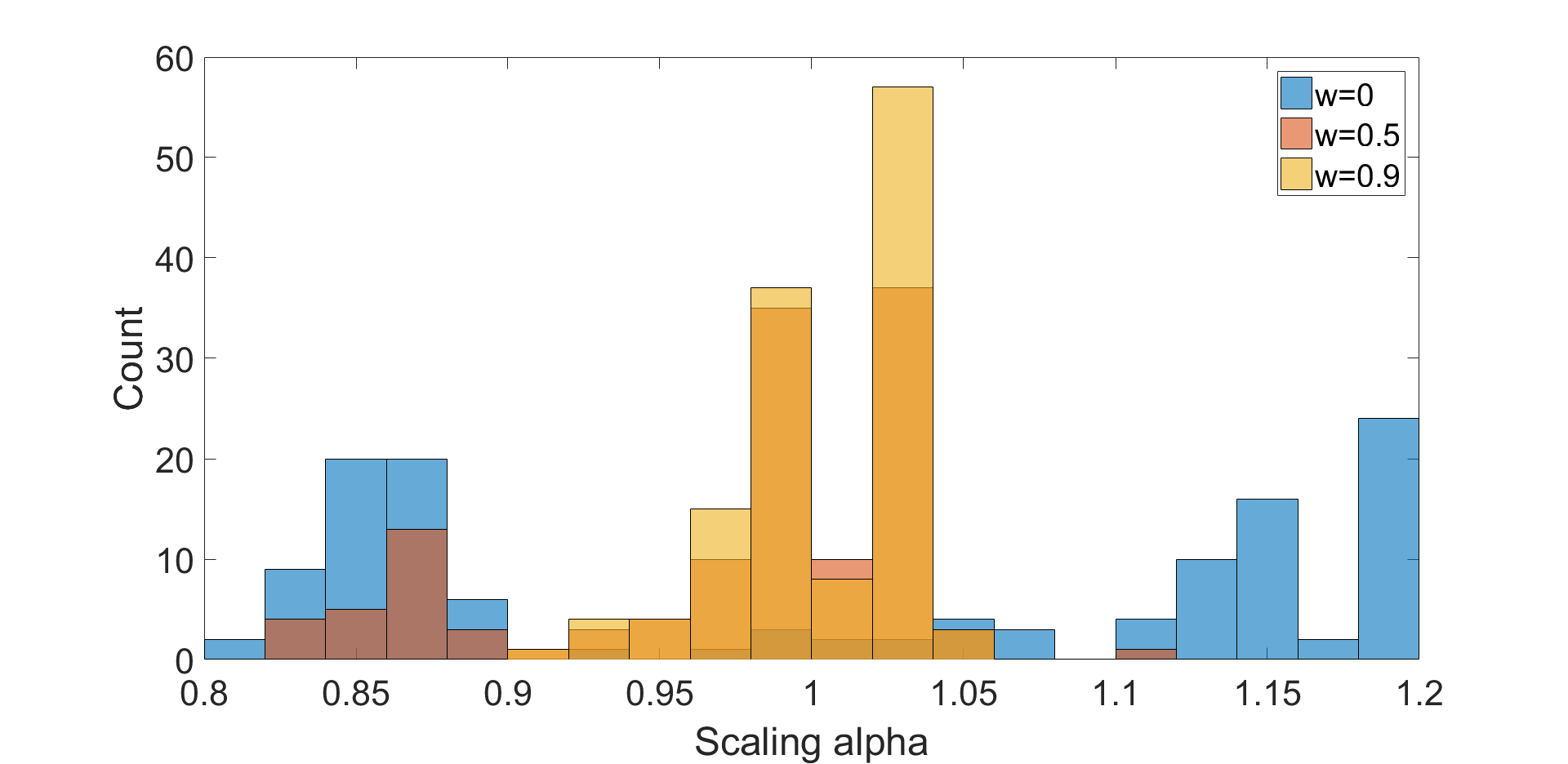}
	\caption{Histogram of the $\alpha$ values for three weights, 0 (blue), 0.5 (red) and 0.9 (yellow).}
	\label{fig:histogram-alpha}       
\end{figure}

For all non-domestic customers, we record their scaling factor $\alpha$ (used in Eq. \eqref{eq:cost-function}) for different weighting parameter values. Note that by definition, $\alpha=1$ when $w=1$, i.e. the SA buddying. We search for relationships between $\alpha$ and the customers' mean daily demand or the proportion of non-domestic customers on the feeder. No such correlation was found to exist. However, $\alpha$ seems to heavily depend on the weight $w$ in Eq. \eqref{eq:cost-function}. In Figure \ref{fig:histogram-alpha}, we plot the histogram of the $\alpha$ values of the non-domestic customers for three values of the weight parameter, $w \in \{0, 0.5, 0.9\}$. We observe that for $w = 0$, $\alpha$ tends to take values on the two extreme sides, as the second term in the cost function Eq. \eqref{eq:cost-function} contributes nothing, therefore there is no suppression of $\alpha$. As $w$ increases, the second term contributes more significantly restricting $\alpha$ to values close to 1. The distribution of $\alpha$ for small $w$ might indicate that the estimated mean daily demand of the non-domestic customers is not accurate, as the algorithms favours values of $\alpha$ away from the unit. However, this is an open question that requires further investigation in future studies.

\section{Feeders with only non-domestic customers}
\label{sec:nondomestic-feeders}

We now consider the feeders which are purely non-domestic since it allows us to investigate the buddying in more detail. In Table \ref{tbl:non-domestic-customers}, we present a summary of the six feeders with only non-domestic properties. The first column is feeder number, second column shows the type of customers on the feeder. Third and fourth columns represent the actual and estimated mean daily demand of the feeder and property respectively. The actual is derived from the actual feeder readings, while the estimated mean daily demand is extracted from the quarterly meter readings (QMR).

\begin{table}[]
	\centering
	\begin{tabular}{|c|l|r|r|}
		\hline
		Feeder Number & Non-domestic Type                                                                                                                                & Actual (kWh) & Estimated (kWh)                                                        \\
		\hline \hline
		1  & \begin{tabular}[c]{@{}l@{}}Community centre,\\  School\end{tabular}                                                                        & 186.01       & \begin{tabular}[c]{@{}l@{}}153.01 \\   71.43\end{tabular}             \\
		\hline
		2  & School                                                                                                                                           & 116.82       & 119.87                                                                 \\
		\hline
		3  & Community centre                                                                                                                                 & 56.96        & 49.80                                                                  \\
		\hline
		4  & Landlord lighting supply                                                                                                                         & 290.99       & 8.49                                                                   \\
		\hline
		5  & Landlord lighting supply & 1797.13      & 2.29                                                                   \\
		\hline
		6  & \begin{tabular}[c]{@{}l@{}}Industrial high load factor,\\   Industrial high load factor,\\ Industrial low  load factor\end{tabular} & 1407.80      & \begin{tabular}[c]{@{}l@{}}39.04 \\   141.20 \\   39.04 \end{tabular} \\ 
		\hline		
	\end{tabular}
	\caption{Summary of feeder characteristics with only non-domestic customers.}
	\label{tbl:non-domestic-customers}
\end{table}

The first thing to note is that the actual mean daily and the daily estimated in the QMR are not always accurate (in fact the sum of the Estimated should be close to the actual). The estimated mean daily demand for feeders 1 to 3 is relatively accurate, but feeders 4, 5 and 6 are quite poor. This is likely because the connection information is incorrect (within the DNOs databases), or the QMR are based on estimates, etc. For example, feeder 5 has been assigned a landlord lighting for an office block but the profile does not resemble this since it has very little overnight demand, see Figure \ref{fig:mean-weekly-demand}. Feeder 6 appears to have a high turnover of businesses and hence it is difficult to keep track of what business was operating when the QMR data was collected (2013). 

\subsection{Feeder profile characteristics}
\label{sec:profile-characteristics}

\begin{figure}
	\centering
	\includegraphics[scale=0.65]{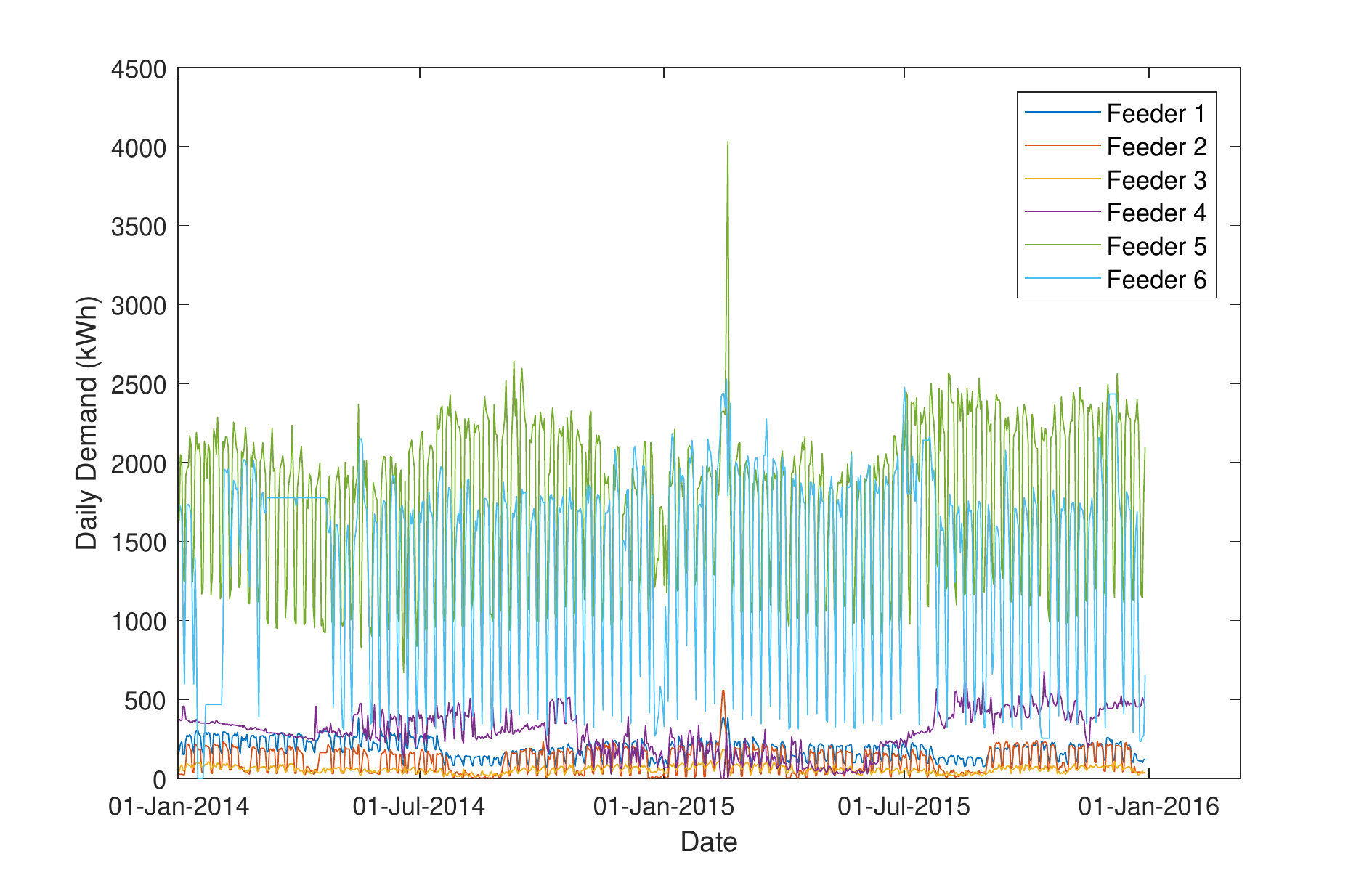}
	\caption{Total daily demand profile of the six feeders for a two-year period. We note that there are errors in readings for feeder 6 within the first 100 days.}
	\label{fig:two-year-profiles}       
\end{figure}

First, we consider the daily usage of the feeders over the years 2014 and 2015. This is shown in Figure \ref{fig:two-year-profiles}. From Figure \ref{fig:two-year-profiles}, we observe the following. Firstly, compared to domestic customers the seasonal increase at winter time does not appear \cite{Haben2016}. Also, the range of behaviours of non-domestic customers is obviously very different. Feeder 6, in particular, seems to take a wide range of daily demands, clearly linked to their weekly pattern. Feeders 1, 2 and 3 have low demand usage for several periods in the following days in 2014: trial days 202 (21st July) to 247 (4th Sep.) which correspond to the summer period and trial days 354 (20th Dec.) to 369 (4th Jan. 2015) which is the Christmas period. Similarly, the same periods in 2015 are also identified. This is unsurprising since feeders 1, 2 and 3 are actually connected to schools/community centres which may be less busy in the school holiday periods. The size of demand on Feeders 4 and 5 suggests these are not simply landlord supplies. Finally, we note that there seems to be a regime change in feeder 4 from around trial day 550 (14th December 2014) onwards. This could be due to new uses of the building.

\begin{figure}
	\centering
	\includegraphics[scale=0.6]{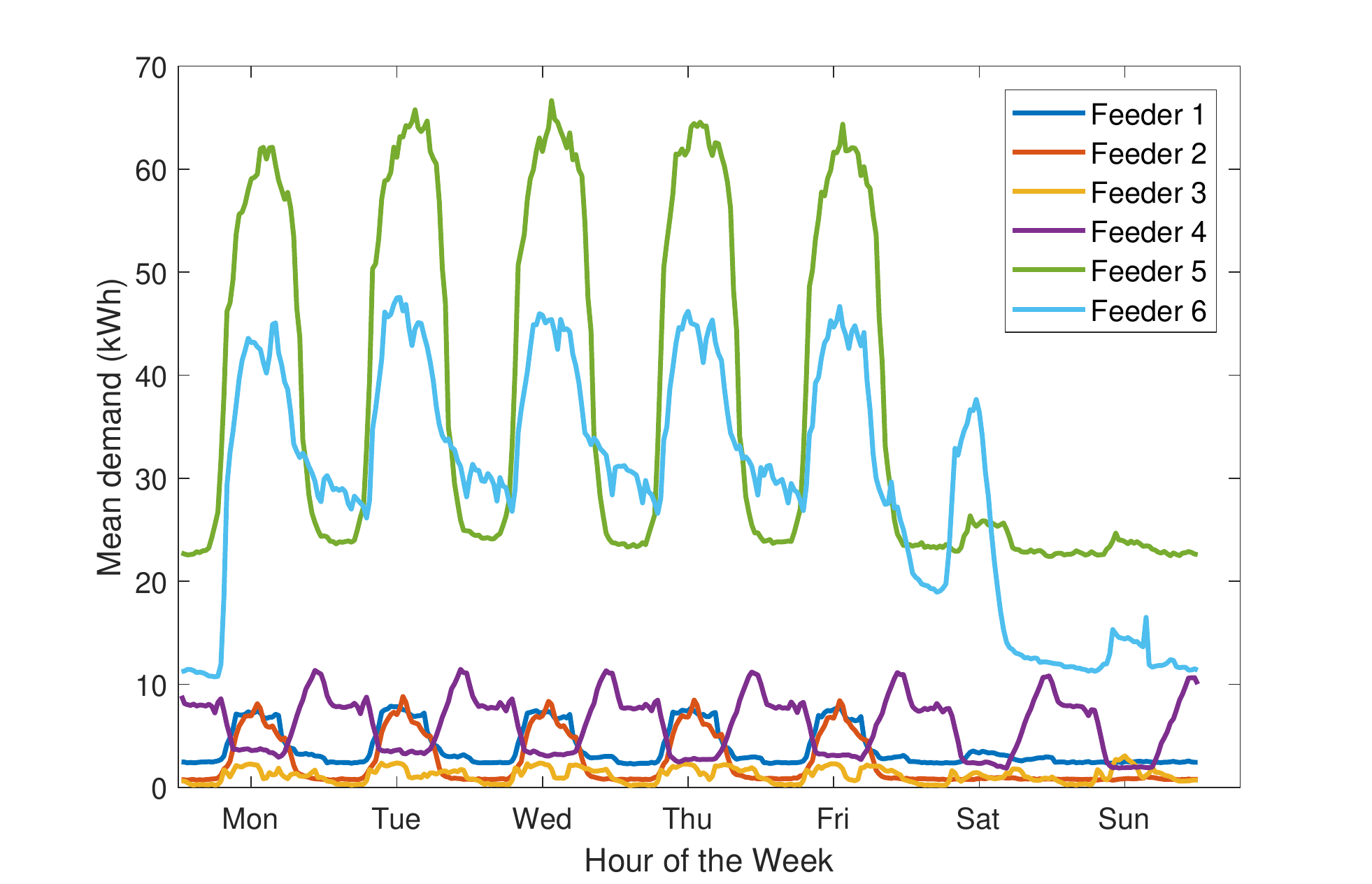}
	\caption{Mean weekly profile for the six feeders.}
	\label{fig:mean-weekly-demand}       
\end{figure}

To investigate more details of these feeders we also consider their average weekly profile, shown in Figure \ref{fig:mean-weekly-demand}.

The daily demands have been ordered starting from Monday through to Sunday. The weekly patterns are more obvious for feeders 5 and 6. Indeed, it appears that feeder 5 is connected to a commercial which is non-operational at the weekends, which contributes further evidence that this has been mislabelled as a landlord lighting supply, and on feeder 6 there is some demand on Saturday but with a shorter working period and lower demand on Sunday.

The profile of feeder 4 seems to confirm that this customer is landlord lighting because the highest demand occurs around the hours of 5-6 p.m. until around 9a.m. Feeders 1 and 2 have low weekend demands (feeder 1 has a small amount of demand on Saturday) which is perhaps as you would expect for community centres and school demands. 

Some of the conclusions from this analysis are the following; i) There can be a large mismatch between the daily mean usage and those demands as recorded through the quarterly meter readings, this could be because the connectivity information is inaccurate and/or the readings themselves are inaccurate (due to estimated and not actual quarterly readings). ii) There are clear weekly patterns in the data which can inform what sort of commercial customers are connected. iii) There are features in the data which are also uncommon to residential only networks. In particular, for the networks and customers covered in this study, there are many differences between the weekend vs weekday behaviours, annual seasonality is not that prominent, and also there are potential changes in regime that can occur. This could be due to the fact that non-domestic properties are more susceptible to churns\footnote{The term \textit{churn} refers to a customer leaving a supplier.} and that new technologies could have an effect on baseline/typical demands. Such inconsistencies will have an impact on the potential accuracies of the methods as we will show in the section \ref{sec:comparison-windebut}. 

\subsection{Comparison with the standardised profiles}
\label{sec:comparison-windebut}

Using standardised profiles from network modelling tools \cite{Windebut} we can construct the normalised profiles for a variety of commercial customer which we utilise in the buddying. Here, we analyze the characteristics of such profiles, perform buddying and try to draw some conclusions against the actual data that we use.

\begin{figure}
	\centering
	\includegraphics[width=\textwidth]{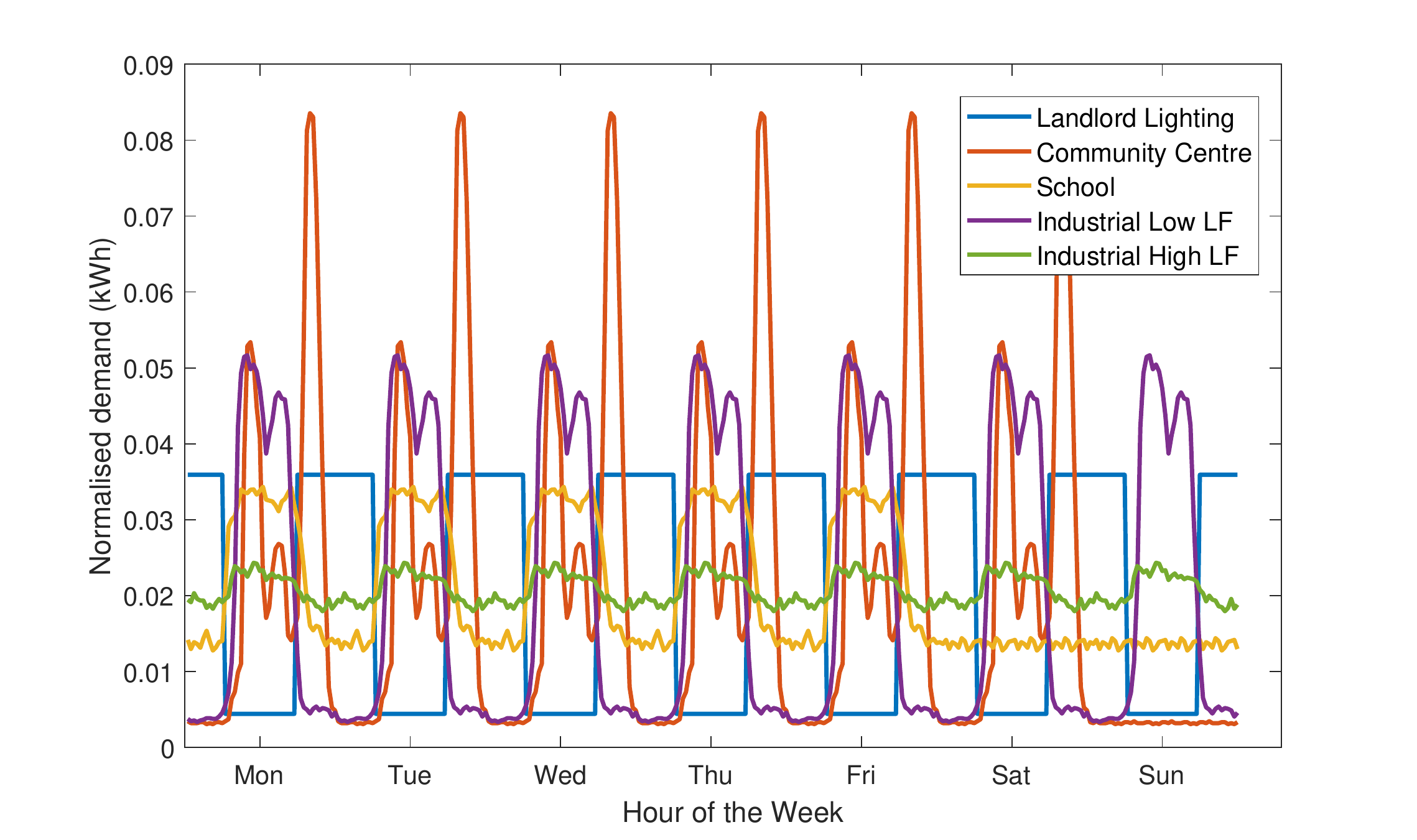}
	\caption{Representative normalised weekly profile for the types of the non-domestic customers on the six feeders.}
	\label{fig:normalised-profiles}       
\end{figure}

Using information from the DNO database we can locate the non-domestic customers on a particular feeder. We then use public information to identify the type of business, e.g. supermarket, library, etc. After extracting a standardised profiles for typical operational and non-operational day for the business, we then must extend the profiles across the year based on assumptions or readily available information.  For example, with a school, a sensible use of the profiles is to use the weekend profiles for the summer/Easter breaks, etc., since this is likely a closer match with the true profile. The profile is then normalised so they have mean daily usage of 1 unit. The normalised profiles for the main non-domestic customers considered in this section (see Table \ref{tbl:non-domestic-customers}) are shown in Figure \ref{fig:normalised-profiles}.

The landlord lighting seems to be consistent with overnight charging seen in the feeder profile. However, they turn off earlier (7a.m.) and turn on at 6p.m. compared to 9a.m. and 5p.m. for the real feeder. Note that these times would likely move around throughout the year. The community centre has the same general shape of the real feeder data but without the larger afternoon spike relative to the daytime. Clearly the assumption that the industrial high/low load factor customers work on the weekend is incorrect for modelling feeder 6. The school, however, looks quite consistent with the demand on feeder 2. The accuracy of the profiles is clearly dependent on a number of factors, in particular having accurate information of the type of business and their operational hours. This can be quite a labour intensive activity and hence maintaining an up-to-date database is essential for practical application of the methods proposed.

We will now apply three buddying methods, including the two versions of the simple buddy. Note we do not buddy feeders with multiple commercial customers (feeders 1 and 6) to simplify the analysis.
Focusing on single-commercial feeders does not restrict the applicability of the method but simplifies the analysis and presentation of the effect of estimated and actual mean daily demand on the methods.
 We test over the same period using the same 8 weeks as with the full buddying study outlined in Section \ref{sec:buddying}. When scaling the normalised profiles, we use three methods, (i) scaling by the actual mean daily demand (i.e. the simple buddy assuming perfect quarterly meter reading information), (ii) scaling by the estimated mean daily demand (the usual simple buddy outlined in Section \ref{sec:buddying}), and (iii) optimal, i.e. letting the GA decide what is the best scaling $\alpha$ when $w=0$, see Eq. \eqref{eq:cost-function}.

We compare the RMAE errors between these three methods in Table \ref{tbl:rame-errors-feeders2-5}.

\begin{table}[]
	\centering
	\begin{tabular}{|l|r|r|r|}
		\hline
		RMAE ($10^{-5}$)     & Estimated & Actual & Optimal \\ \hline \hline
		Feeder 2 & 4.45      & 4.41   & 4.54    \\ \hline
		Feeder 3 & 5.04      & 5.31   & 4.99    \\ \hline
		Feeder 4 & 5.53      & 5.78   & 4.83    \\ \hline
		Feeder 5 & 5.7       & 5.75   & 3.81    \\ \hline
	\end{tabular}
	\caption{RMAE errors of non-domestic profiles of feeders 2 to 5 for three methods.}
	\label{tbl:rame-errors-feeders2-5}
\end{table}

The first thing we notice is that the Optimal profile gives the best match over the yearly test period for all feeders except feeder 2. The biggest improvement is in the feeder 5 with over 30\% improvement on both the estimated and actual scalings. Feeder 4 has over 10\% improvement using the optimal scaling compared to the two simple buddy scalings. There is not much difference in the feeder 2 scores (i.e. the school customer). Comparing the two simple methods (i) and (ii), we can see that they are similar in most cases but using the estimated daily mean demand performs slightly better than using the actual for three cases. Some of the differences in accuracy can be explained by considering the scalings used. The scaling for each method are presented in Table \ref{tbl:scaling-feeders2-5}.

\begin{table}[]
	\centering
	\begin{tabular}{|l|r|r|r|}
		\hline
				 & Estimated 	& Actual 		& Optimal \\ \hline \hline
		Feeder 2 & 119.87      	& 116.82   		& 84.86    \\ \hline
		Feeder 3 & 49.80      	& 56.96   		& 34.67    \\ \hline
		Feeder 4 & 8.49      	& 290.90   		& 216.44    \\ \hline
		Feeder 5 & 2.29       	& 1797.13   	& 652.79    \\ \hline
	\end{tabular}
	\caption{Scaling of the non-domestic profiles for three methods}
	\label{tbl:scaling-feeders2-5}
\end{table}

All the methods performed equally well for feeder 2 and from Table 4 we can see that each method (especially the estimated and actual simple methods) gives similar scalings. Feeder 5 is poorly estimated by all methods, this is because the profile used for the matching was a landlord lighting profile and the feeder was clearly a daytime usage customer. The optimal method performed much better here simply because the operational period of the landlord lighting period was matched to the non-operational period of the actual profile, minimizing the errors as much as possible.

\begin{figure}
	\centering
	\includegraphics[scale=0.55]{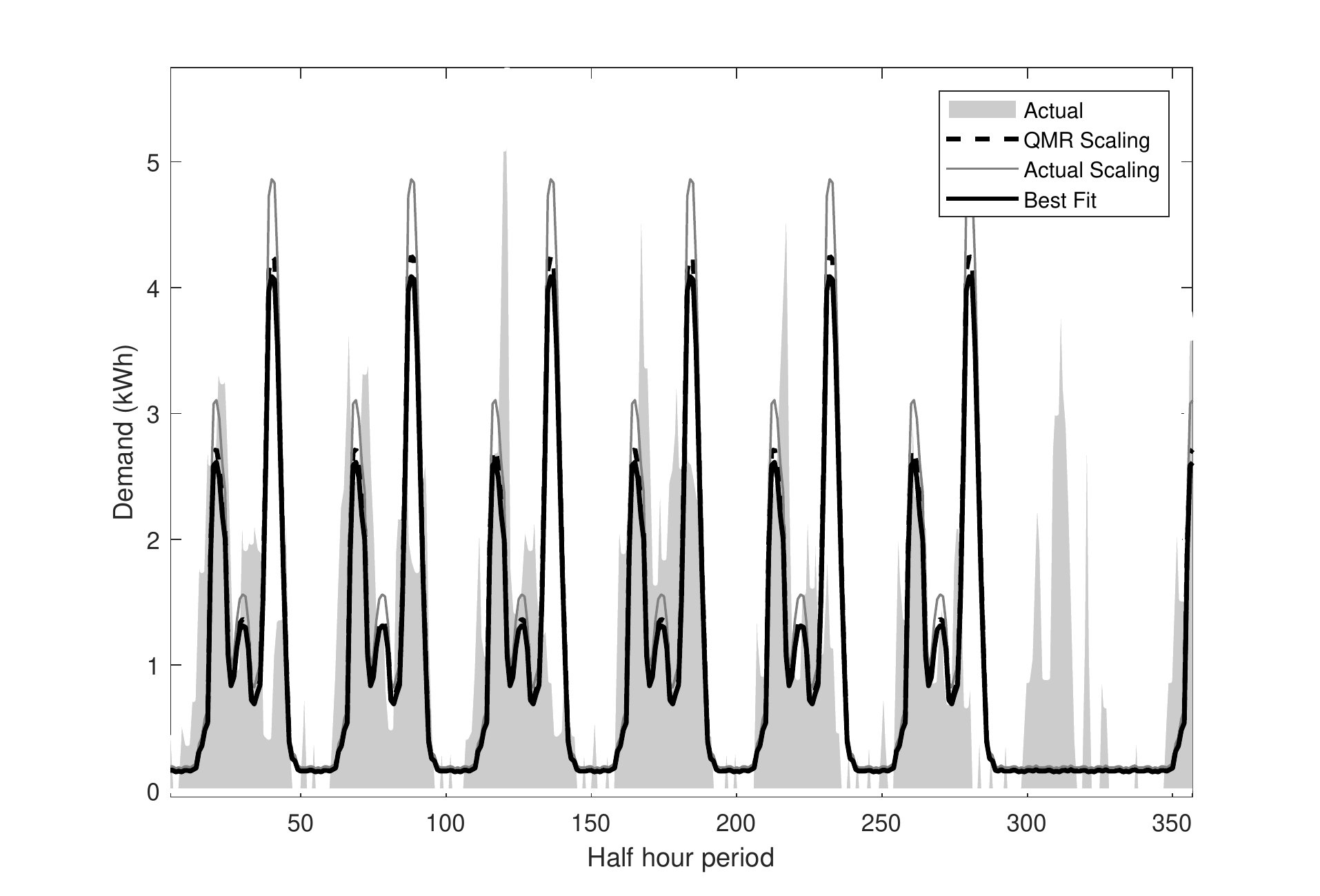}
	\caption{Actual and model demand of feeder 3 for seven consecutive days}
	\label{fig:actual-vs-model-commercial}       
\end{figure}

The matches are accurate for all methods for feeder 3 except for on the weekend, see Figure \ref{fig:actual-vs-model-commercial}, because we assumed no demand on the Sunday when creating our normalised profiles. However, the other days are well modelled and thus the scores and the optimal scalings are similar for all methods. Feeder 4 is modelled poorly since the assumed customer connected to this feeder is not consistent with the actual profile and there appears to be more than a single landlord lighting demand on this feeder.

This section shows then that no matter how good the match, we are limited in accuracy if the assumed customer and their normalised profile are inconsistent with the actual feeder. Hence feeders with non-domestic customers are much more sensitive to errors in accurate quarterly meter readings, in reporting of the type of business and even knowledge of their operational and non-operational hours. We have seen, especially with feeders 2 and 3 that accurate modelling is possible but accurate records are more necessary when commercial customers are connected to a feeder. 

\section{Confidence}
\label{sec:confidence}

The buddying method populates a network with realistic profiles for all customers without the requirement of large-scale monitoring. However, the demand of an LV network is volatile and further techniques are required to understand the available headroom of the network for planning and development. For example, in the UK, network operators have used well-established procedures, such as the after diversity maximum demand (ADMD) and ACE49 \cite{ENA1981}, to estimate the demand and design LV networks. Models of estimating the load uncertainty, particularly the uncertainty of the peaks, of an LV network are essential for planning and management. In this section we consider alternative methods for generating confidence bounds at the feeder level based on limited network modelling.

\subsection{Methodology}
\label{sec:confidence-methodology}

In this section, we present two methodologies for modelling feeder demand uncertainty. In the first scenario, a simple method that assumes no substation monitoring data relies on bootstrapping the demand from the monitored customers (and the standard normalised profiles) to develop a range of profiles which we can then model as an empirical distribution and select confidence bounds. The second method requires substation monitoring data and uses a quantile regression model to forecast the confidence levels. The latter method is more accurate, but more computationally expensive and requires data and/or monitoring of the LV substations/feeders.

To assess the accuracy of the forecasts we will consider a standard measure, the continuous ranked probability score (CRPS) \cite{Gneiting2007}, calculated according to the quantiles at 90\% and 10\%. We consider a normalised CRPS where we divide by the mean half hourly demand over the chosen year period. That way we can compare feeders of different typical magnitudes. The models will be fit 90\% and 10\% confidence for the date of the last year of the data (we have 552 days starting from 20th March 2014) and we calculate the errors for the dates 23rd September 2014 to 22nd September 2015.

\subsubsection{Bootstrapping}
\label{sec:bootstrapping}

Assume we have a feeder with $M_{\text{dom}}$ domestic customers, of which $M_1$ are profile class 1 and $M_2$ are profile class 2, and $M_{\text{com}}$ non-domestic customers. For each bootstrap \cite{Efron1993}, we sample, with replacement, from our monitored domestic customers $M_1$ profile class 1 customers and $M_2$ profile class 2 customers. For each of the $M_{\text{com}}$ non-domestic customers, we sample the scaling factor which will then be applied to the normalised profiles. The scaling will be randomly sampled from (i) a uniform distribution, (ii) a Gaussian distribution. In the former case, the range of the uniform distribution is on the closed interval $[0.8, 1.2] \times U_j$, where $U_j$ is the mean daily usage for that customer (similar to the discussion in the GA in Section \ref{sec:buddying}). In the latter case, the Gaussian distribution has mean equal to the estimated mean daily demand, $U_j$, and standard deviation $\sigma = 20\mu/196$, which is found so that 95\% of the demands are in the range $[0.8, 1.2] \times U_j$ of the customer type $j$. We run 1500 bootstraps from which we empirically select the 90\% and 10\% quantiles.

\subsubsection{Quantile Regression}
\label{sec:quantile-regression}

For this method, we fit a model to the quantiles of choice to the historical data \cite{Koenker2001}. There are a large range of models that could be used \cite{Hong2016}, but here we use a simple seasonal model where each half hour of the day is modelled with a separate equation of the form

\begin{equation}
	\hat{L}_d^{\tau}(\mathbf{\alpha^{\tau}}) = \alpha_0^{\tau} + \alpha_1^{\tau} d + \alpha_2^{\tau} S_d + \alpha_3^{\tau} \hat{S}_d + \sum_{p=1}^{P}\left( b_p^{\tau} \sin \left(\frac{2 \pi p d}{365}\right) + c_p^{\tau} \cos \left(\frac{2 \pi p d}{365}\right) \right),
	\label{eq:quantile-regression}
\end{equation}
where $d$ is the day of the year (starting with $d=1$ for the first day of the dataset), $S_d$ is a dummy variable indicating whether the day is a Saturday or not, and $\hat{S}_d$ is a Sunday identifying dummy variable. Hence, the model only considers a simple linear trend, annual seasonality and weekend effects. The chosen quantile $\tau$ (either the 10\% or 90\% quantile), is found by minimizing the pinball function, defined as 
\begin{equation}
	\rho_{\tau}(z) = |z \cdot (\tau - \mathbf{1}_{(z<0)})|,
\end{equation}
where $\mathbf{1}_{(z<0)}$ is the indicator function, 
over the dataset (see \cite{Haben2016b} for more details).

\subsection{Results}
\label{sec:confidence-results}

\begin{figure}
	\centering
	\includegraphics[scale=0.6]{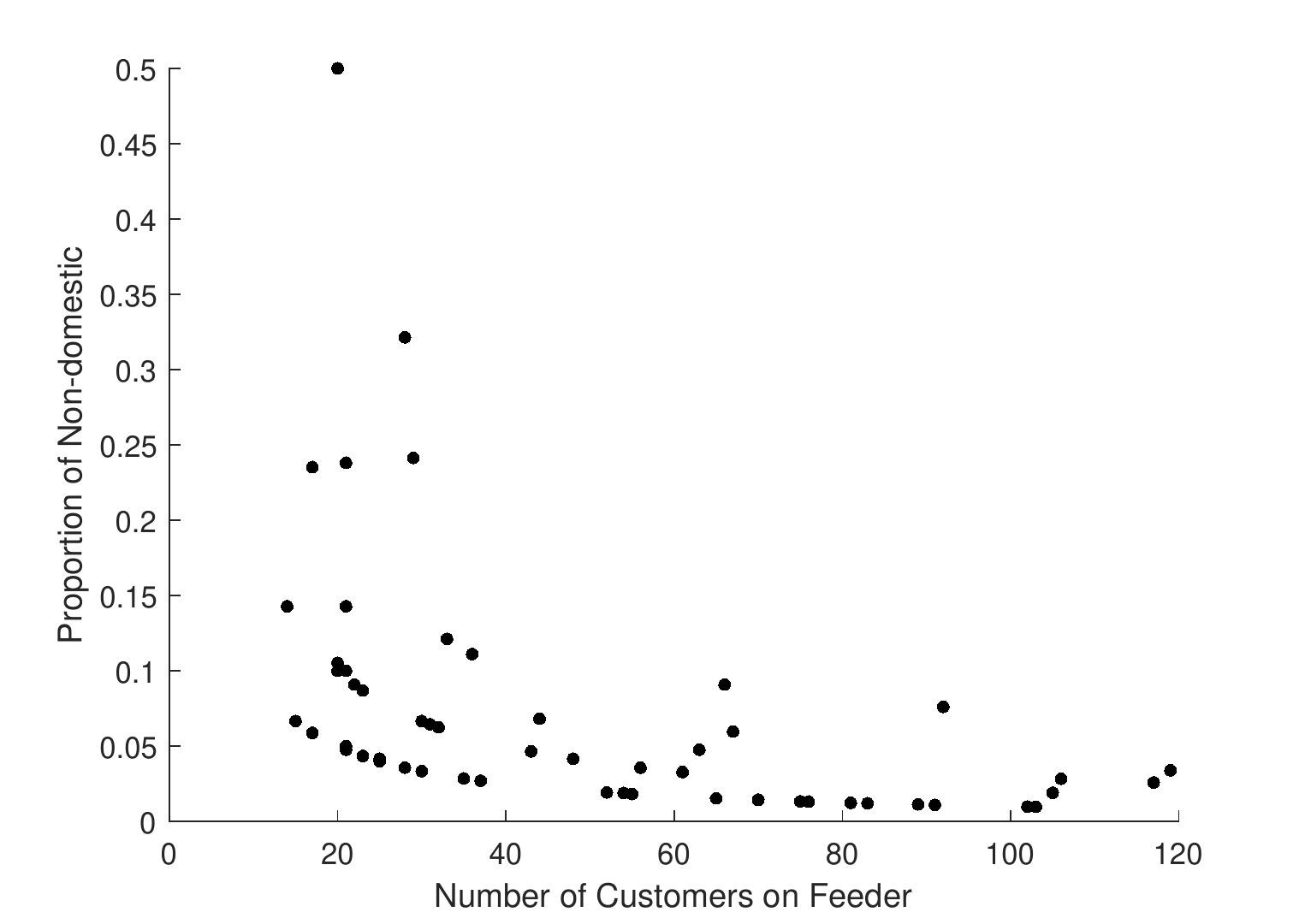}
	\caption{Relationship between the proportion of non-domestic customers versus the number of customers connected to a feeder.}
	\label{fig:prop-non-domestic-vs-feeder-size}       
\end{figure}

There are 58 feeders which have non-domestic customers, whose proportion of non-domestic varies from as small as 1\% of the customers on a feeder to 50\%. Figure \ref{fig:prop-non-domestic-vs-feeder-size} shows the relationship between the proportion of non-domestic and the number of customers connected to the feeder. Clearly, the smaller number of customers the wider ranging proportions (since smaller numbers of customers will have larger effects), but often feeders consist of less than 10\% commercial customers. 

A main question is how different proportions of commercial customers affect the feeder uncertainty modelling of our methods. We split the feeders into four categories for analysis: Those with no commercial, those with a small proportion of commercial (less than 5\%), those with medium proportion (5 -10\%) and those with large ($>10$\%), there are 127 with no commercial, 33 with small proportion, 13 with medium and 12 with large. 

Table \ref{tbl:crps-3-mehtods} shows the average normalised CRPS scores for different types of feeders and for the three confidence methods (quantile regression, bootstrap with uniform distribution and bootstrap with Gaussian distribution).

\begin{table}[]
	\centering
	\begin{tabular}{|l|p{1.6cm}|p{1.6cm}|p{1.6cm}|}
		\hline
		Feeder type                    & Quantile Regression	& Bootstrap Uniform	& Bootstrap Gaussian         \\ \hline \hline
		All                            & 0.123          & 0.403			& 0.403                      \\ \hline
		Domestic only                  & 0.125					& 0.350 		& 0.350                      \\ \hline
		Small proportion of non-domestic  & 0.111					& 0.479			& 0.479                      \\ \hline
		Medium proportion of non-domestic & 0.118					& 0.376			& 0.376                      \\ \hline
		Large proportion of non-domestic  & 0.14					 & 0.773 			& 0.773 										 \\ \hline
	\end{tabular}
	\caption{Normalised CRPS error of the confidence bounds for three methods.}
	\label{tbl:crps-3-mehtods}
\end{table}

Firstly, the substation based confidence is quite consistent no matter what the connectivity in the feeder. The worst errors are in the feeders with largest proportion of non-domestic customers but only about 10\% worse than the domestic only feeders. Secondly, it is clear that there is more inconsistency in the bootstrapping methodology. They are consistently worse than the substation confidence for all types of feeders but are very poor for the feeders with large proportions of non-domestic customers. Feeders with Large proportion of non-domestic have errors over 120\% larger than domestic only feeders. We note that the bootstrap methods will always be identical when considering domestic only feeders since the techniques are identical. 

In addition, both bootstrap methods have very similar results for all feeder types and very large errors for the feeders with large proportions of non-domestic customers. This indicates many sources of potential errors in the bootstrap including (i) the initial poor diversity generated by using the domestic sampling (hence we likely need a larger sample), (ii) the quarterly meter readings are not very accurate, especially for non-domestic customers and (iii) the information about the non-domestic customers is incorrect and hence we have constructed an inaccurate profile for them. Note for all feeders the substation confidence outperforms the Bootstrap methods and is, on average, 3.3 times better. Since the errors are very similar between both bootstrap methods from here on we only consider the uniform version since it is slightly more accurate. 

\begin{figure}
	\centering
	\includegraphics[width=\textwidth]{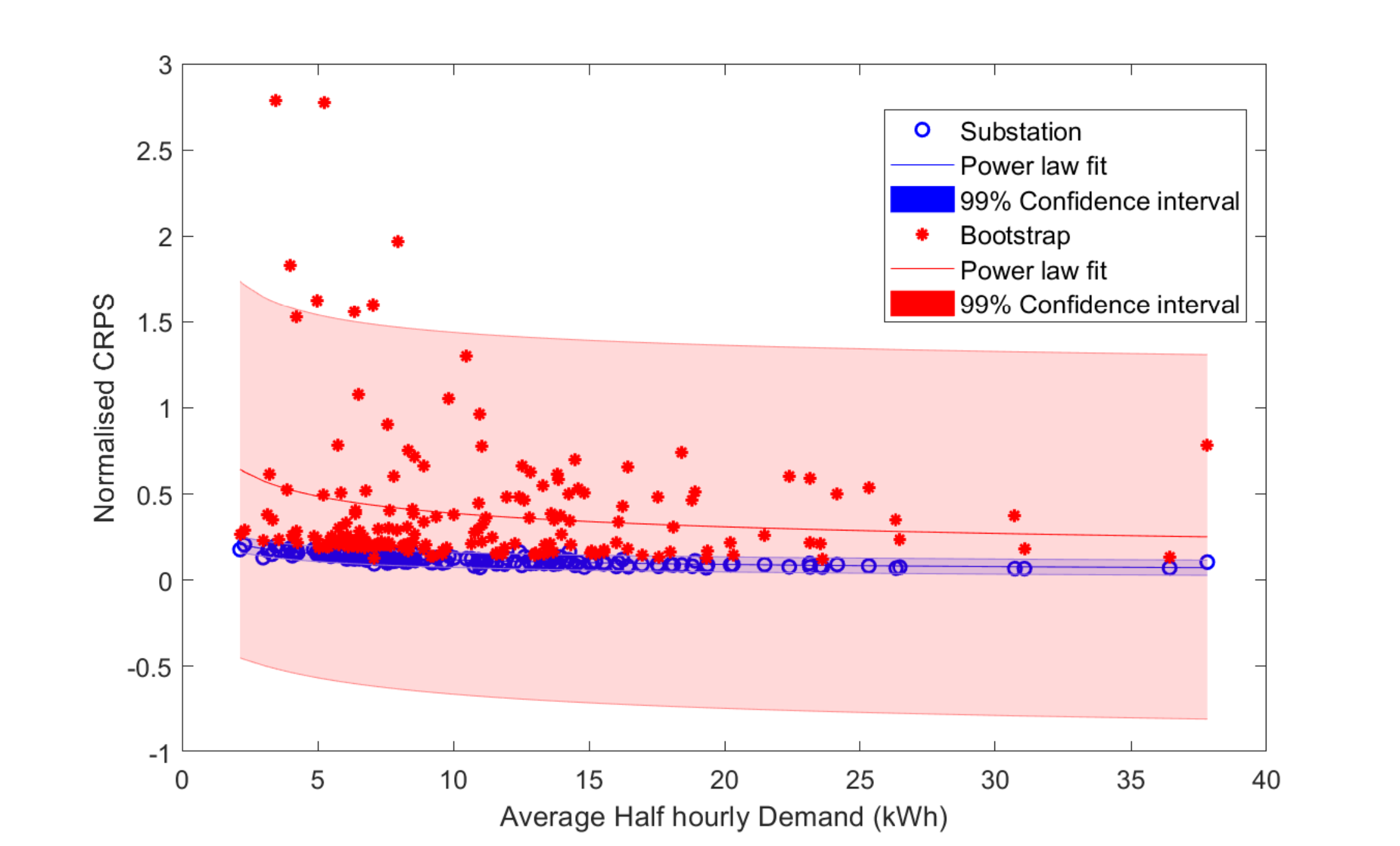}
	\caption{Normalised CRPS as a function of the mean demand of the feeder.}
	\label{fig:crps-vs-avg-demand}       
\end{figure}

Now, we consider the normalised errors as a function of the average half hourly load of the feeder.  We expect and find, see Figure \ref{fig:crps-vs-avg-demand}, that the confidence of the larger feeders is more accurately computed than on smaller feeders. This is because larger feeders are either the result of larger numbers of customers or are the result of larger, and more predictable commercial customers. Hence these larger feeders will have smoother demand and are easier to model.  The plot in Figure \ref{fig:crps-vs-avg-demand} shows the normalised errors for each feeder as well as a power law fit and the 99\% confidence bound.

\begin{figure}
	\centering
	\includegraphics[width=\textwidth]{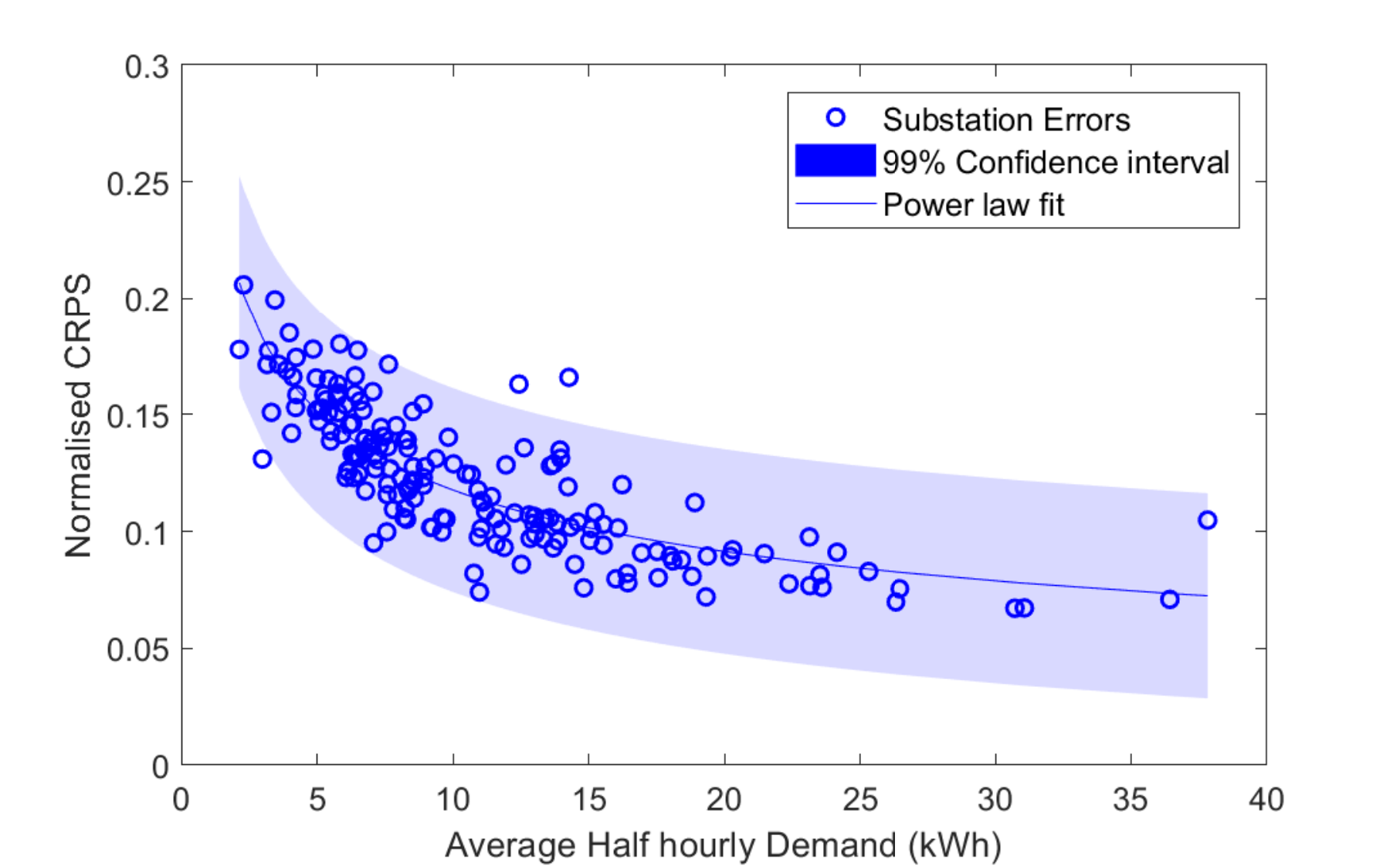}
	\caption{Normalised CRPS as a function of the mean demand of the feeder only for the quantile regression method.}
	\label{fig:crps-vs-avg-demand-qr}       
\end{figure}

Clearly, the fit of the quantile regression method, which is based on the substation monitoring, is tighter and thus more accurately modelled. To highlight this, we plot the quantile regression method on its own in Figure \ref{fig:crps-vs-avg-demand-qr}. The relationship shows a power law fit and more accurate modelling as the size of the feeder (in demand) increases. There are slightly less points for the larger substations so ideally, we would like more of these feeders to improve our confidence in the analysis.

\begin{figure}
	\centering
	\includegraphics[scale=0.6]{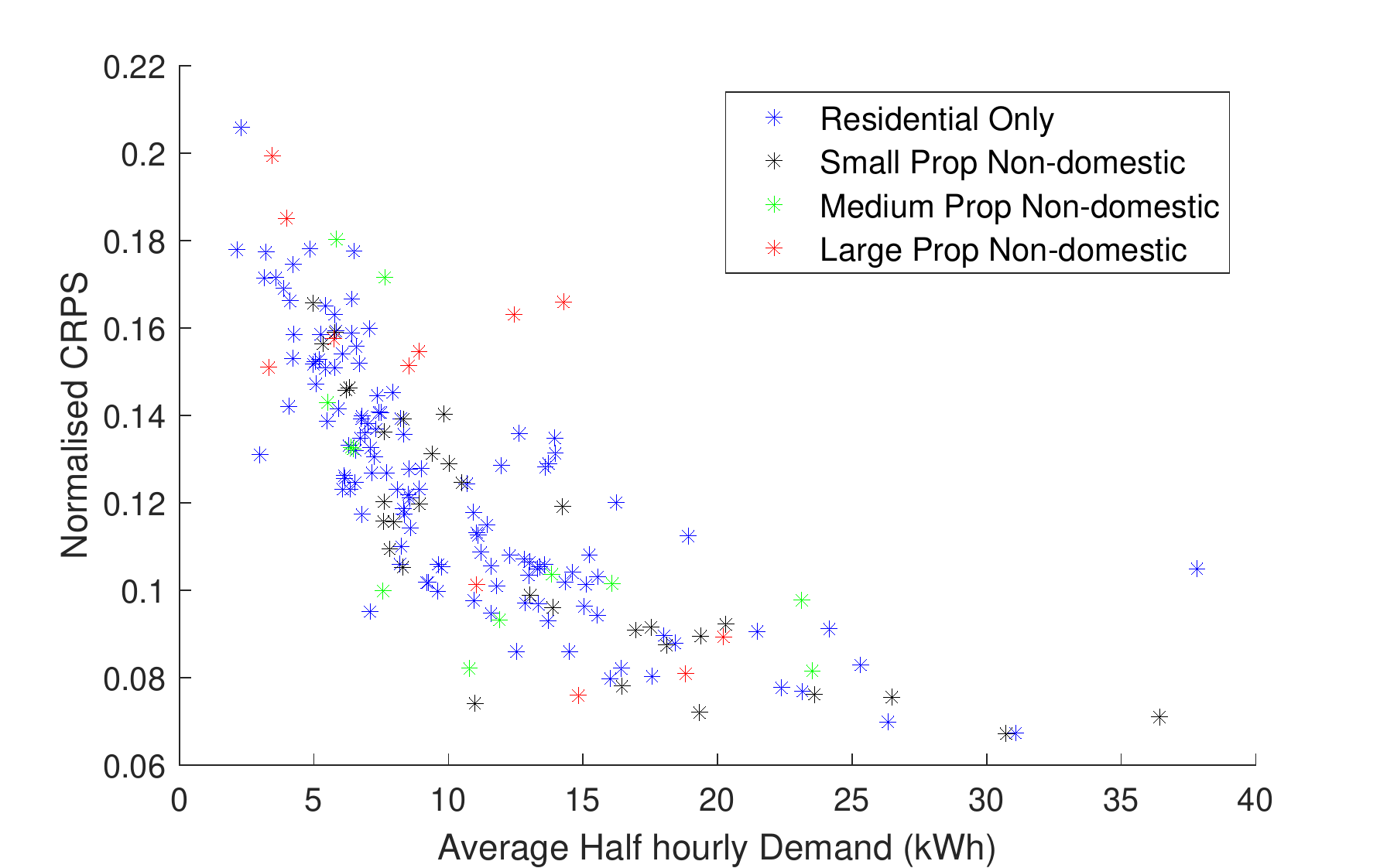}
	\caption{Normalised CRPS as a function of the mean demand of the feeder for the quantile regression method with colour-coded feeder types.}
	\label{fig:crps-vs-avg-demand-qr-coloured}       
\end{figure}

We investigate the effect of the commercial customers in more detail now. Figure \ref{fig:crps-vs-avg-demand-qr-coloured} is similar to Figure \ref{fig:crps-vs-avg-demand-qr}, but we also show the feeders coloured as an indicator of the proportion of non-domestic.

Here we see the proportion of non-domestic customers on the feeder is obviously not a strong factor in determining the accuracy of the confidence estimates. Although the feeders with large proportion of non-domestic customers do deviate a little from the main shape this is only for 2 of the 12 feeders. In addition, the magnitude of demand on a feeder does not necessarily indicate the mix of customers on that feeder, this is highlighted by the small proportion group (black) which are distributed quite evenly as a function of mean demand. Similar conclusions also hold for the bootstrap method with uniform distribution, where there is no significant influence by the mix of customers on a feeder.

\subsection{Examples}
\label{sec:examples}

\begin{figure}
	\centering
	\includegraphics[scale=0.6]{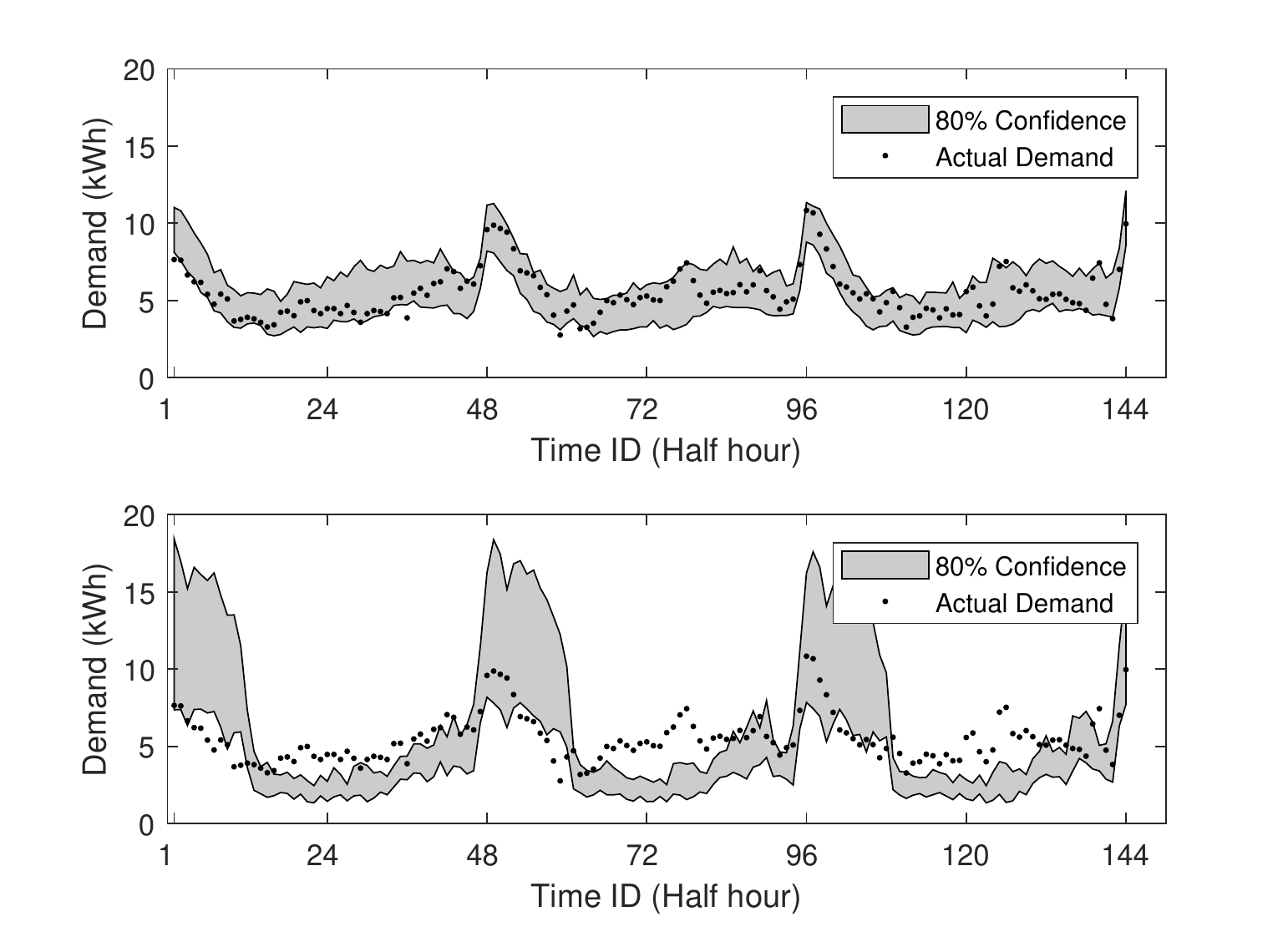}
	\caption{Profile of feeder 2 for three days with confidence bounds for quantile regression method (top) and bootstrap method (bottom).}
	\label{fig:example-feeder2}       
\end{figure}

In this section, we present a few examples of the confidence analysis for three consecutive days starting on 17th of November 2015. Feeder 2, from Section \ref{sec:nondomestic-feeders}, has a school attached to it. Figure \ref{fig:example-feeder2} shows the confidence bounds of the quantile regression method (top) and bootstrap method (bottom). Clearly the substation based confidence is much more accurate than the bootstrap but in both cases the general shape has been captured.

\begin{figure}
	\centering
	\includegraphics[scale=0.6]{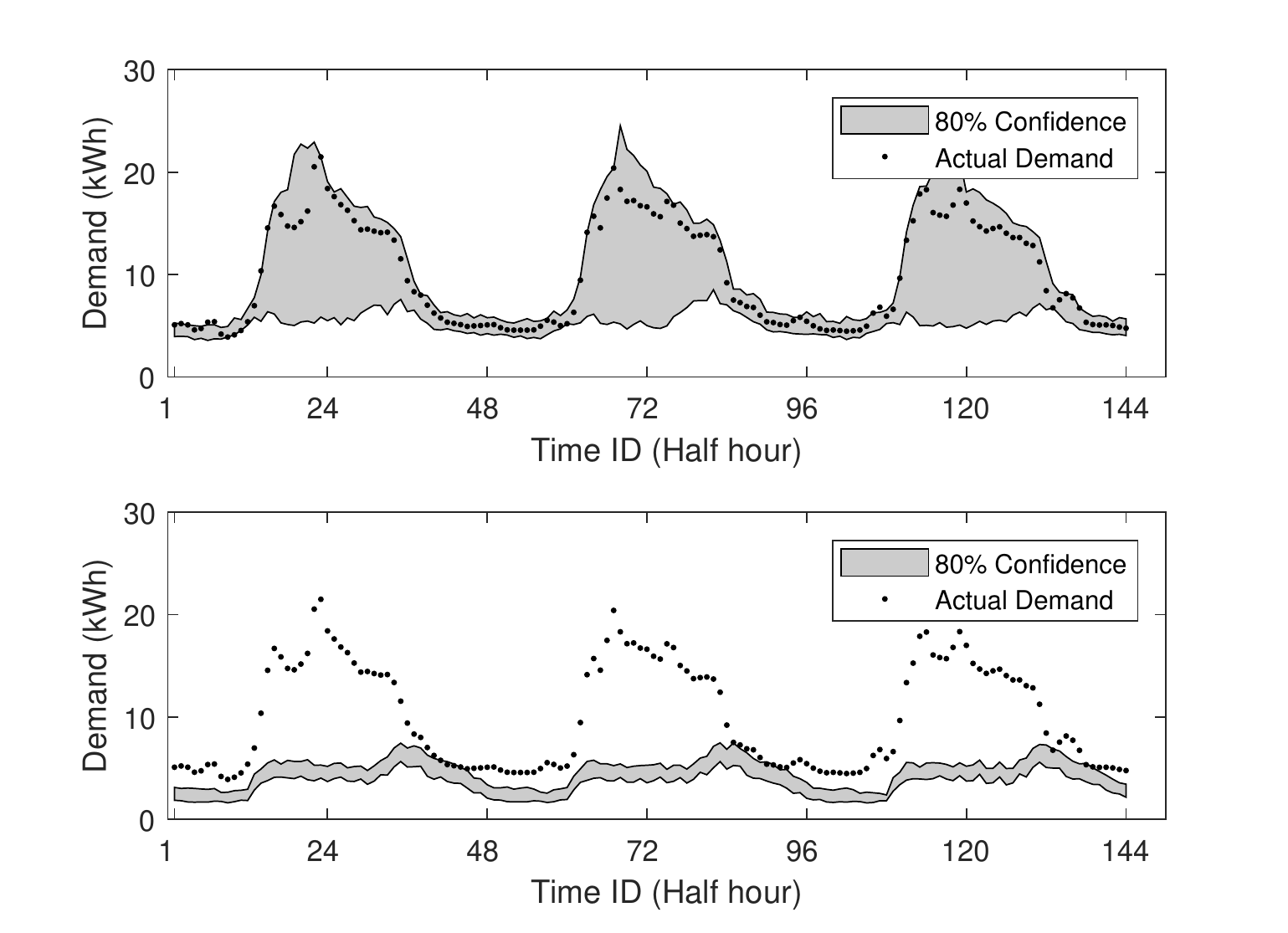}
	\caption{Profile of a feeder with large proportion of non-domestic customers for three days with confidence bounds for quantile regression method (top) and bootstrap method (bottom).}
	\label{fig:example-large-prop}       
\end{figure}

The following feeder, shown in Figure \ref{fig:example-large-prop}, has a large proportion of commercial on a feeder of 21 customers. This time the bootstrap method has completely missed the daytime demand. In this case the mean daily demand records of the commercial customers are likely inaccurate since the feeder is dominated by the day time usage of commercial customers but the bootstrap method is essentially showing a domestic dominated feeder.

\begin{figure}
	\centering
	\includegraphics[scale=0.6]{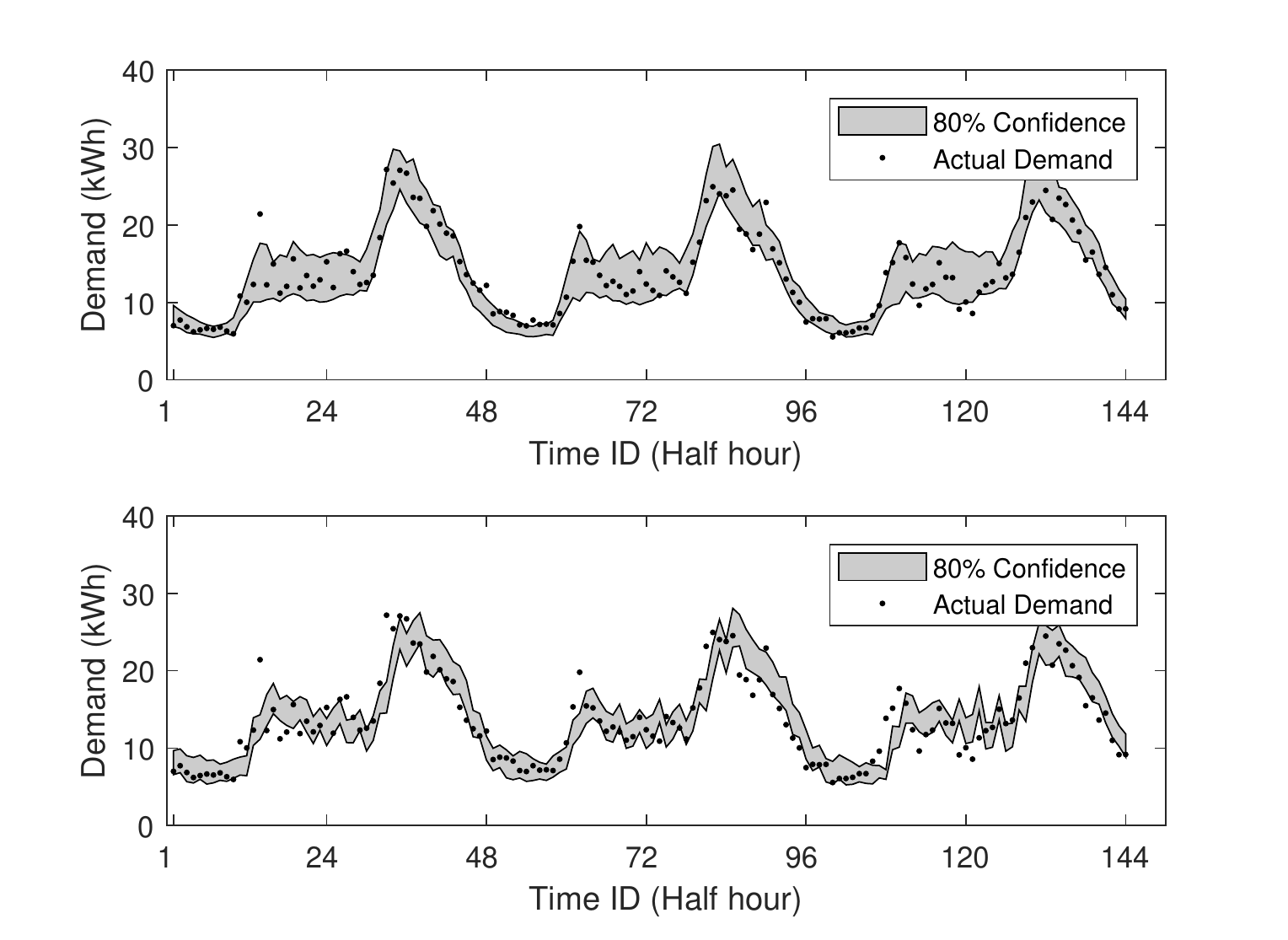}
	\caption{Profile of a feeder which has domestic-only customers for three days with confidence bounds for quantile regression method (top) and bootstrap method (bottom).}
	\label{fig:example-domestic-only}       
\end{figure}

The feeder in Figure \ref{fig:example-domestic-only} is purely domestic with 146 customers. Both methods have performed well, likely due to the large numbers of customers generating more regular demand which is easier to estimate. In addition, this figure illustrates that the bootstrap methods may be better suited to domestic only feeders until better quarterly meter reading information is available.

\section{Conclusion}
\label{sec:conclusion}

The spread of low carbon technologies is drastically changing energy consumption and generation patterns of LV networks. Despite the wide roll-out of smart meters, this data might not be available to DNOs and if it is, it might be expensive. In this study, we extended a recently introduced method in \cite{Giasemidis2016}, called buddying, to non-domestic customers. Due to the absence of monitoring data for non-domestic customers, the annual profiles were created from standard weekly profiles from network design tools. We showed that modelling feeders with non-domestic customers is, on average, less accurate than modelling purely domestic feeders. The reason is three-fold: (i) the mean daily demand estimate of some non-commercial customers might be inaccurate and an order of magnitude off the real value, (ii) the connectivity information of the feeder might be inaccurate, due to faulty reporting and/or churn, and (iii) the annual profiles of the non-domestic customers were constructed by simply extending the corresponding weekly standard profiles considering only bank and school holidays. Future refined models must also consider seasonality and trend effects in the demand.  

Due to the volatile nature of LV networks, confidence bounds on the load are necessary for modelling and operational purposes. We introduced two techniques that estimate confidence bounds, one that uses substation monitoring data and the other based on bootstrapping from the monitoring data. We demonstrated their performance and searched for factors, such as the proportion of non-domestic customers, that might be indicative of the method’s performance.

The main application of these buddying profiles is to serve as baseline profiles into power flow analysis tools without needing large amounts of monitored data to be available. Hence such models could assist a network operator in designing, planning and managing their LV infrastructures. In addition, the baseline profiles can be combined with different future scenarios of low carbon technology to simulate future networks \cite{Hattam2017}. Although the focus of the proposed methods, buddying and uncertainty intervals, is on the smart grid sector, both can be extended to other sectors, e.g. gas, heating, leading to an integrated solution to smart energy systems \cite{Lund2017556}.

\section*{Acknowledgements}
We would like to acknowledge the support of Scottish and Southern Electricity Networks (SSEN) and its personnel during the collaboration of the New Thames Valley Vision Project (SSET203 New Thames Valley Vision), funded via Ofgem's Low Carbon Network Fund.

\bibliographystyle{plain}
\bibliography{bibfile}

\end{document}